\documentclass[12pt, draftclsnofoot, onecolumn]{IEEEtran}
\usepackage{lettrine}
\usepackage{epsfig}
\usepackage{footnote}
\usepackage{float}
\usepackage{amssymb}
\usepackage{amsthm}
\usepackage{amsmath,mathabx}
\usepackage{mathtools}
\DeclarePairedDelimiter{\ceil}{\lceil}{\rceil}
\usepackage{color}
\usepackage{amsfonts}
\usepackage{epstopdf}
\usepackage{breqn}
\usepackage{tabularx,tabulary}
\usepackage[font=footnotesize]{caption}
\usepackage{graphicx}
\usepackage{caption}
\usepackage{subcaption}
\usepackage{wrapfig}
\usepackage{dblfloatfix}    
\usepackage{bbm}
\usepackage[noend]{algpseudocode}
\makeatletter
\newcommand{\removelatexerror}{\let\@latex@error\@gobble}
\def\BState{\State\hskip-\ALG@thistlm}
\makeatother
\usepackage{atbegshi}
\AtBeginDocument{\AtBeginShipoutNext{\AtBeginShipoutDiscard}}

\begin{document}
\title{F4Tele: FSO for Data Center Network Management and Packet Telemetry}

\author{Amer~AlGhadhban}

\thanks{Authors are with College of Engineering, Electrical Engineering Dept. at University of Hail, KSA.}
\maketitle
\thispagestyle{empty}
\pagestyle{empty}



\begin{abstract}
The proliferation of bandwidth-hungry applications and services forces datacenter (DC) administrators to optimize the utilization of available resources.  Precisely, the network share of management traffic has grown significantly because DC networks are becoming more sophisticated and require a massive amount of data for efficient debugging and troubleshooting~\cite{everflow, pingmesh, plank}. Accordingly, we use free space optics communication (FSO) with wavelength division multiplexing (WDM) technology to build a flexible yet high-performance logical network responsible for management traffic. The FSO-WDM can provide reconfigurable multi-terabit topology over line-of-sight (LoS) links. Due to space and processing capacity reasons, we can not offer direct connections from every data rack to the network management racks. Alternatively, the data racks are grouped together as each group is serviced for a duration of time matches its average arrival-rate. Since the data racks showed different arrival-rates, the hotspot racks are allocated with longer service time. The evaluation results show that F4Tele carried out high throughput close to the expensive solution (benchmark). 
\end{abstract}
\IEEEpeerreviewmaketitle
\begin{IEEEkeywords} 
Wavelength routing, wavelength assignment, network management, free space optical communications, delay analysis, lightpath provisioning, wireless data centers, wireless optical communications.  
\end{IEEEkeywords}
\newpage
\section{Introduction}
 
The cloud service providers (e.g., Microsoft Azure and IBM SmartCloud), mobile operators (e.g., China Mobile and Verizon), electronic commerce companies (e.g., Amazon and Alibaba) and content delivery services (e.g., Google CDN and Akami) strive to enhance their data center network (DCN) performance to cope with the demands of an ever-growing market and inclusively its bandwidth-intensive applications such as the proliferation of e-Business, e-Government, smart cities, and big data. These applications continuously exchange a vast volume of data among an enormous number of servers.

These applications need to operate on powerful servers connected by a multipath network of high bisection bandwidth and availability. Building such a network requires elbow grease and dedication:  optimizing computational and communication resources for maximal power-savings and minimal cost, and at the same time committing to communication and security policies. A network installation is often the easy part in the journey of a DC construction. The real challenge is how to efficiently manage the DCN to keep it running correctly and fulfill the service level agreement. This challenge reaches its momentum when the managed business heavily depends on its online presence; the value of network management techniques cannot be overstated. 
Typically, DCNs are designed for high utilization, and even subtle performance degradation or short-term failure can lead to high losses. A network management solution enables easy detection and identification of network issues before they turn into a dilemma. Hence, ensuring minimum maintenance time and the least impact on DC productivity. 

On the other hand, the DCN encounters various factors of uncertainty, such as diversity of faults, dynamic workloads, and wide-spectrum applications, creating a troublesome environment.  Understanding and debugging faults in DCNs is challenging because failures have different shapes and similar impacts. For example, some packets may experience long round-trip times, but it is unclear which network components are responsible. Also, a packet drop has multiple root causes: hardware errors, software bugs, misconfigurations, or congestion. The congestion itself occurs due to transient reasons, or high traffic loads, or permanent reasons, such as partial interface failure. Thereby, debugging faults in DCNs requires capturing a large number of packets and sending them to a powerful unit for analysis~\cite{pingmesh,everflow,plank,flowradar}.

It is impossible to mirror a large number of packets without introducing an overhead on the data network because it takes too much bandwidth to transmit the captured packet. For example, Pingmesh~\cite{pingmesh} is a network management framework designed by Microsoft for latency measurements and analysis. It gathers large volume, ($\approx$50 terabytes), of network measurement data per day. Clearly, this amount introduces long queuing delay and consecutive packet dropping when it is forwarded through the same network of data packets. Similarly, EverFlow~\cite{everflow} collects large amount of tracing packets for network failure debugging and troubleshooting that force it to dedicate multiple servers and switches to fulfill the high-processing demand to analyze the collected data. On the other hand, Planck~\cite{plank} utilizes the port-mirroring feature in DC switches to measure the links statistics. The solution mirrors as much as it can the buffered packets of all outgoing to a centralized collector. Such solution causes extra processing overhead and cabling from Top-of-Rack (ToR) switches to the centralized collector. The number of racks in normal DCs ($\approx$10K) makes such techniques particularly difficult unless mirrored packets are being transmitted through the data network or a new network is dedicated only for the mirrored packets. FlowRadar~\cite{flowradar} encodes the statistics of every flow in the switch memory, the encoded statistics are updated with every packet and exported to a central analyzer per 10ms. Although it entails significant modifications for every network switch to maintain low communication and processing overhead, it needs 2.3Gbps per switch to send the collected statistics, and encoded flowsets to the NMS. Other researchers have introduced alternative solutions at the expense of the necessary performance figures, (e.g., accuracy, and speed), and features (e.g., routing loops and blackhole detection)~\cite{flight, opensketch, pathdump, ltsm, lightfdg}.


The captured packets in these schemes are transmitted along with the real data through the same network resources that cause an unpleasant impact on network performance, especially if the monitored incident is for critical applications, users, or security. Therefore, DCN introduces unique challenges which necessitate new solutions and different procedures than the ones used in a conventional network. Hence, in this work, we present F4Tele, which uses a high speed yet flexible topology FSO to transmit the captured packets to NMS racks. Recently, Ciaramella et al.~\cite{Ciaramella2009128} achieved a total of 1.28 Tbps speed in an outdoor experiment of 212 meters distance by using WDM-FSO link of 32 wavelengths (32$\times$40 Gbps). When using wireless FSO links between the server-racks and the NMS-rack(s) three features are enabled: re-configurable topologies, high link capacity, and low cabling complexity and maintenance overhead. The previous researches~\cite{firefly, projector,  Celik2018LightFD, lightfdg} used FSO technology to transmit data traffic. However, they suffered from several challenges (e.g., switching speed, FSO alignment, and LoS congested horizon). Alternatively, the management traffic has some characteristics (e.g., joint direction, constant destination, scattered traffic sources, and less congested horizon) that make it convenient with FSO positives and not affected by its negatives. 

Although this solution looks dull, we have encountered multiple implementation challenges — transceiver spatial challenges, traffic load diversity, and FSO beam alignment and spinning overhead. Intuitively, a normal DCN has thousands of ToR switches, and the physical dimension and processing capacity of a rack are not enough to install or process thousands of transceivers to communicate with every data rack. We exploit recent DC findings in solving these challenges. The researchers~\cite{facebook,projector} found that the DC communication demonstrated bias distribution where a few numbers of racks (hotspots) are the destination of about 80\% of the traffic. Thereby, the DC racks can be divided into mainly two classes according to the arrival-rate: hotspot and non-hotspot.  In this work, the hotspot racks are grouped together, and similarly, the non-hotspot racks. Also, the FSO topology reconfigurability has been used to optimize the service-time assignment strategy by assigning high capacity and long service time for the hot-spots group, and the remaining time and capacity are used for the non-hotspot group. Moreover, other than rotating the transceivers, we use multiple ceiling reflectors and switchable mirrors to overcome the beam alignment challenge and delay.
\subsection{Paper Objectives and Achievements}
Main objectives and contributions of this work can be summarized as follows:
 \begin{itemize}  
\item[$\bullet$]
Recently the DC network devices, including servers, exchange large volumes of management traffic~\cite{everflow,pingmesh,flowradar,plank}. This traffic is highly significant, and it has a direct impact on several valuable services such as failure debugging and troubleshooting platforms as well as traffic engineering systems. F4Tele exerts full attention to this traffic by utilizing a flexible yet high-speed technology to transmit it away from data traffic. Other than forwarding this large amount of significant traffic through the data network, forcing it to compete with data traffic, F4Tele reaps the benefits of FSO emerging technology to build a flexible and high-performance network dedicated only for network management traffic.  
\item[$\bullet$] 
The spatial space and processing capacity of NMS racks are not enough to serve the FSO beams in one-time. The DC has thousands of racks, which means thousands of FSO transceivers on top of the NMS racks. Instead, F4Tele divides the data racks into multiple sets of equal size as much as possible (not larger than NMS capacity). The assembling of set members is performed according to the arrival-rate of their NMS traffic. Every data racks that showed equal NMS arrival-rate $\lambda_{m}$ are grouped together in the same set. F4Tele allocates service-time intervals for those sets matching their $\lambda_{m}$. The FSO beams of the same sets are constantly pointed toward the same reflector in the ceiling that is connected to a control system to control their service-time.
\item[$\bullet$] 
In order to avoid the delay and the overhead of adjusting and spinning the FSO beams and mirrors, F4Tele employs soft measures to simplify its structure and achieves its goal. In F4Tele, the FSO beam switching overhead migrated from the edge to a centralized system. Hence, every FSO beam of a single data rack is pointed to a reflector on the ceiling. The F4Tele structure has multiple reflectors, one for every set of data racks where a few numbers of sets are served at a time. Besides this, the ceiling reflector is supported with a transferable mirror where it has a switchable background to control its transparency. The state of these mirrors is controlled by a centralized microswitch that is connected to an SDN controller. The microswitch is preconfigured to switch on and off the mirror according to the arrival-rate of the set. The mirror blocks the beam of unserved sets and unblocks the beam of the under service set(s). The blocking and unblocking are according to the instruction from the microswitch. In this case, F4Tele doesn't need to change the direction of FSO transceivers or the mirrors during its operations. 
\item[$\bullet$] 
The F4Tele has been evaluated by implementing it in Minine-HiFi~\cite{mininet} and POX controller~\cite{pox}. The evaluation results show that F4Tele carried out high throughput close to the expensive solution (benchmark) during TCP and UDP traffic and for different arrival-rates.  
\end{itemize}
\subsection{Paper Organization}
The rest of the paper is organized as follows: The background and related work is presented in Section \ref{sec::relatedwork}. The problem statement is explained in Section~\ref{sec::statement}. After that, the solution architecture, rotation-time as well as service-period assignment, and delay analysis are discussed in Section~\ref{sec::solution}. Then, the implementation and performance evaluations are presented in Section~\ref{sec:evaluation}. Finally, conclusions are drawn in Section~\ref{sec::conclusion}.  
 \section{Related Work}\label{sec::relatedwork}
Wireless DCNs are mostly studied in the realm of physical topology design which generally tries to establish connectivity among the racks. For instance, the authors of ProjecToR~\cite{projector} leveraged digital micromirror devices and disco-balls to speedup the switching of FSO links. The digital micromirror device can direct FSO beams toward tens of thousands directions, while it needs 12$\mu$s to switch between these directions. In \cite{Celik2018Design, Celik2019Design, Celik2018LightFD, lightfdg} the authors utilized FSO-WDM and traffic grooming to augment the DCN bandwidth and exploit the FSO-WDM agility to build two isolated virtual topologies: one for delay-sensitive flows and another for throughput-hungry flows. The results have demonstrated that the proposed method provides superior performance in both throughput and flow completion times.  In \cite{lightfdg} the authors used a light-weight and fast flow classification methods to steer every flow class to its allocated virtual-topology. 

Existing network management schemes can be categorized into in-host, in-network~\cite{opensketch} and centralized~\cite{plank, everflow, flight}. The  in-host solutions need to modify the kernel network stack of the DCN hosts and hence require higher privileges that cannot always be granted. Also, the operation itself (i.e., loading a kernel module in every tenants) is  troublesome administrative overhead and vulnerable to misconfiguration failures. In the in-network schemes the flow information along with their counters are maintained at the switch memory, and the arithmetic operations are performed internally at the switch hardware (e.g., HashPipe~\cite{heavyhitter}, and OpenSketch~\cite{opensketch}), software (e.g., Software Defined Counter~\cite{sdc}),or otherwise is exposed to a centralized entity (e.g., FlowRadar~\cite{flowradar}, and NetFlow). One of the main challenges of in-network mechanism is the limited resources of network switches (e.g., memory, processing delay, and power consumption). As a result, the researchers introduced the hybrid solution~\cite{pathdump, opensketch, pingmesh,everflow} where some of the operations are migrated into cheaper hardware~\cite{opensketch}, edge~\cite{pathdump}, or into a centralized entity~\cite{flowradar}. For example, OpenSketch~\cite{opensketch}  performs the flow classifications at the fast, small and expensive TCAM memory, while the counters are maintained at a larger, slower, and cheaper memory (SRAM). On the other hand, some solutions (e.g., FlowRadar~\cite{flowradar} and NetFlow) the flow-statistics are exported from the switches to a centralized entity either periodically or with every cache miss. Another challenge in the in-network scheme is altering the existing network switches in the DCN or to use a programmable hardware (e.g., P4) to run their algorithms. 

Although the centralized schemes~\cite{plank,everflow,pingmesh, pathdump,flight} are similar to the in-network hybrid schemes, there are no serious modifications to the network switches or end host devices. It employs the available network features (e.g., packet sampling, port mirroring, or statistical polling, OpenFlow features) to collect network statistics and forward them to a centralized collector which is usually programmed to perform the flow classification functions. However, the existing centralized schemes have some limitations including high monitoring overhead, capturing large volume of data, periodic probes, latency and/or accuracy.

\begin{figure}[t!]
\begin{center}
\includegraphics[width=0.9 \columnwidth]{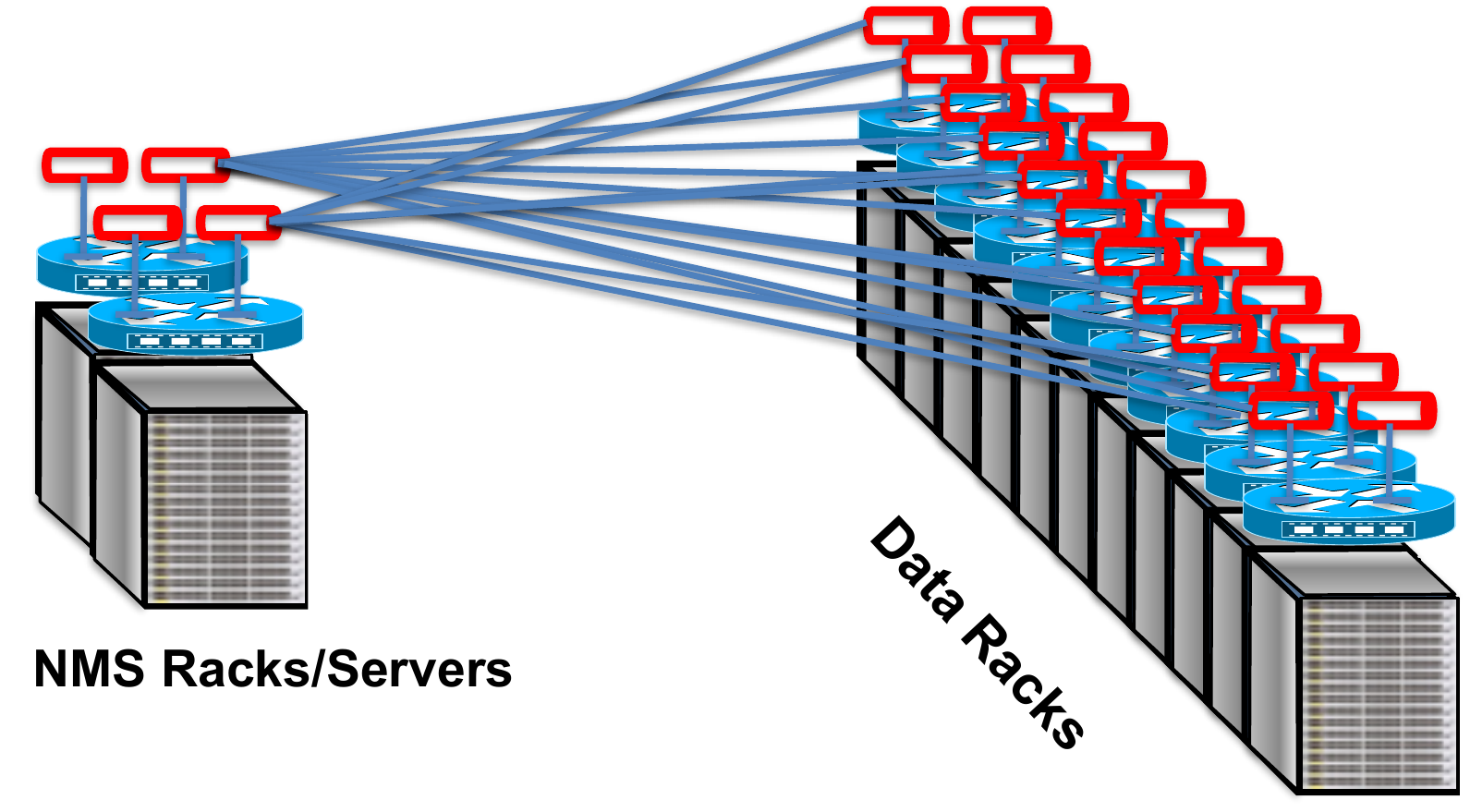}
\caption{Toy example of NMS oversubscription challenge: All the data racks directing their FSO links toward the NMS racks.}
\label{fig:nms-oversub}
\end{center}
\end{figure}

\section{Problem Statement}\label{sec::statement}

The network under consideration has Clos topology, e.g., leaf-spine or fat-tree, with multipath links between rack-pairs which offer high bisection bandwidth and availability. The network has thousands of racks, and each rack has tens of physical servers. The number of data server racks, $R_{D}$, is $N$, while the network has multiple network management servers (NMS) grouped in $U$ numbers of racks, $R_{M}$. The NMS servers have network management applications and storage, e.g., syslogs, SNMP, NTP, TACACS, and network analyzers. 

DC has limited resources where the researches are in full swing to raise their positive utilization. Intuitively, routing these large volumes of control/log traffic through the data network causes direct and indirect negative consequences on DC performance, such as delays, packet droppings, congestion, and creation of bottleneck node in distributed systems. On the other hand, the network researchers compete to introduce efficient packet scheduling, congestion control algorithms, e.g., deadline-aware scheduling, and load balancing schemes to deliver the data packets at the right time and avert passive influences. The straightforward solution is to build a completely new network for NMS communications and avoid routing them via the data network. Unfortunately, a new network means extra administrative overhead and IT operating expenses that include additional network devices, extra staff, and maintenance time as well as troubleshooting. Contrary, FSO technology offers high speed links, programmable topologies, adaptive channel capacity, and naturally improves the physical layer security. 

Rather than sending the control packets via the data network, in this work, we attempt to build an alternative solution by utilizing the extremely high link capacity and the flexibility of FSO to build a high-speed reconfigurable network between the data racks and the network management racks. However, this structure needs at least one FSO transceiver on top of every data rack and the same number on the racks where NMS servers installed.  

Unfortunately, pointing an FSO link from every rack to the NMS rack is almost impossible due to multiple practical obstacles. Practically, the $R_{M}$s have finite physical space and processing capacity. The physical space is not enough to hold all the FSO transceivers for every $R_{D}$ FSO link. Similarly, the $R_{M}$ switch doesn't have enough processing capacity and a number of ports to serve all $R_{D}$ to  $R_{M}$ FSO links. Additionally, there is a high likelihood of collisions and interference between the large bundle of FSO links whenever they are traveling toward nearly the same point. 
 The Fig.~\ref{fig:nms-oversub} illustrates an example where a large number of racks attempt to direct their FSO links toward a few number of NMS racks.  As an alternative, in this work, we aim at optimizing the FSO based solution while satisfying the quality of service constraints. The reconfigurability feature of FSO is exploited to rotate the maximum afforded of FSO links together between data racks per unit of time. 
\begin{table}[]
\centering
\caption{Notations, parameters, and variables}
\label{tab:given}
\begin{tabularx}{1\linewidth}{l X}
\multicolumn{2}{l}{\textbf{Notations and given parameters:}}                                 \\
\hline
\\
$R_{D}$ & The set of data racks.\\
$R_{M}$ & The set of network management racks.\\
$N$     & Number of data racks. \\
$U$     & Number of network management racks. \\
$P$     & The maximum number of FSO links that can be served by $R_{M}$s  \\
$\tau$     & Rotation time. \\
$HR_{s}$     & The highly utilized $R_{D}$s, defined also $R_{D}^H$. \\
$LR_{s}$     & The low utilized $R_{D}$s, defined also $R_{D}^L$. \\
$K$     & Number of $HR_{s}$ \& $LR_{s}$ sets. \\
$K_{H}$     & Number of $HR$ sets. \\
$K_{L}$     & Number of $LR$ sets. \\
$d$     & The length of the service-period, where $\tau$ = $\sum_{i}^K d_i$. \\
$\lambda$  & The average flow arrival-rates. \\
$\mathcal{Z}_{k}$ & The number of packets waiting in the FSO link of a $R_{D_k}$ set, where $k$ is the index of the $R_{D}$ set. \\
$\mathcal{Z}_{v}$ & The number of vacation intervals. \\
$\overline{\mathcal{W}}_{k}$ & The average waiting-time of $R_{D}$ set $k$. \\ 
$Pr_{i}$ & The probability that the switches of $R_{D}$ set $i$ is under service\\
$\mathcal{D}_{k}$ & The waiting-time of $R_{D}$ set $k$. Its value ranges from about $d$ to $\tau_{max}$.\\ 
$\mathcal{K}_{L}$ & The complete set of $R_{D}^L$ sets.\\ 

\end{tabularx}
\hrule
\end{table}
\section{Our Solution F4Tele}\label{sec::solution}
In this section, we present our solution F4Tele. As a first step, we will discuss its design specifications and solution architecture. Second, we discuss how the rotation-time, $\tau$, is optimally allocated. Finally, we present the analytical expressions for the control packet transmission time delay $T$ that mainly depends on $\tau$.  

\subsection{Solution Architecture}
According to recent studies of DCN, the racks, as explained above, can be divided into highly-utilized racks (HRs) and low-utilized racks (LRs). The racks from the same class are grouped together into almost equal size sets, where the set size is $\leq$ P. The number of sets in the system is $K$ and $K$ = $\ceil[\big]{\frac{N}{P}}$ sets. The FSO beam is pointed toward a reflector, e.g., diamond mirror, installed on the ceiling to reflect the incoming beams to the right $R_{M}$ transceivers. In our design, each reflector has a special type of mirror where its transparency is switchable. The number of reflectors is equal to the number of  sets $K$, and the $R_{D}$-to-reflector beams are permanent. Each set of $R_{D}$ steers their beams toward one reflector. These reflectors are controlled by a programmable microswitch, e.g., a Raspberry pi device with an OpenFlow protocol. The function of the microswitch is to change the transparency state of the reflectors.

The challenge is that the FSO links of all $R_{D}$ cannot communicate at the same time with the $R_{M}$ while the swiveling of the FSO link gears, transceiver or mirror, is undesirable because it entails robust structure and careful adjustments as well as it causes unpleasant delay. Consequently, F4Tele is designed with care to avoid the need for swiveling. At first, the programmable microswitch is programmed to switch the vertical mirror "ON" and "OFF" with a preconfigured time interval that is consistent with the utilization level of the connected set. When the mirrors are transparent "ON" the related set of $R_{D}$ connects with the $R_{M}$. Moreover, the microswitch is connected to a controlling unit (e.g., SDN controller) to provide a trustworthy programmability feature which is necessary for future enhancements and modifications, e.g., to regulate the rotation speed up or down.

The other challenge is that the control-packets of the detached $R_{D}$ racks are dropped during $\tau$. Even though in our design, the FSO links of $R_{D}$ would not  be terminated during the switchover, we attempt to introduce a high-fidelity design able to reduce the number of packet loses. The easy solution is by enlarging the memory of interface buffer. However, the FSO buffer tuning is expected to cause degradation of network performance. Because, the borrowed share of memory will be taken from switch-mate interfaces. To mitigate this challenge with minimal overhead, F4Tele leverages the buffer of the backup interface. The ceiling mirrors reflect the FSO beam back toward the backup FSO transceiver creating a routing loop between the primary and backup transceivers. Moreover, the backup interface is preconfigured with a low forwarding rate to keep the incoming packets buffered as much as possible until the rotation is complete. .   

\begin{figure}[t!]
\begin{center}
\includegraphics[width=1 \columnwidth]{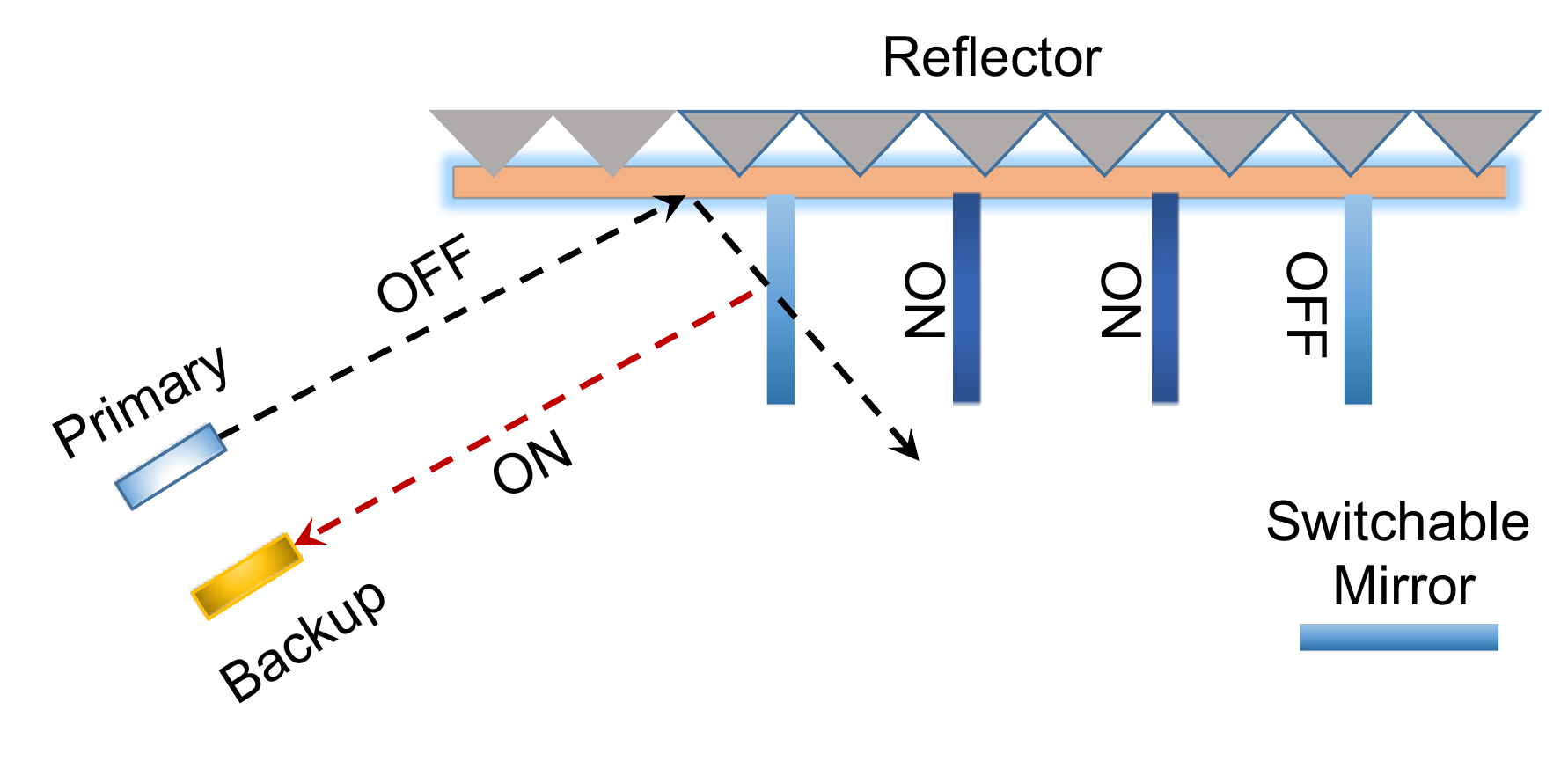}
\caption{Explanatory diagram to show how the FSO beam is routed from primary to backup interface through the switchable mirror (vertical).}
\label{fig:microswitch}
\end{center}
\end{figure}

\begin{figure}[b!]
\begin{center}
\includegraphics[width=1 \columnwidth]{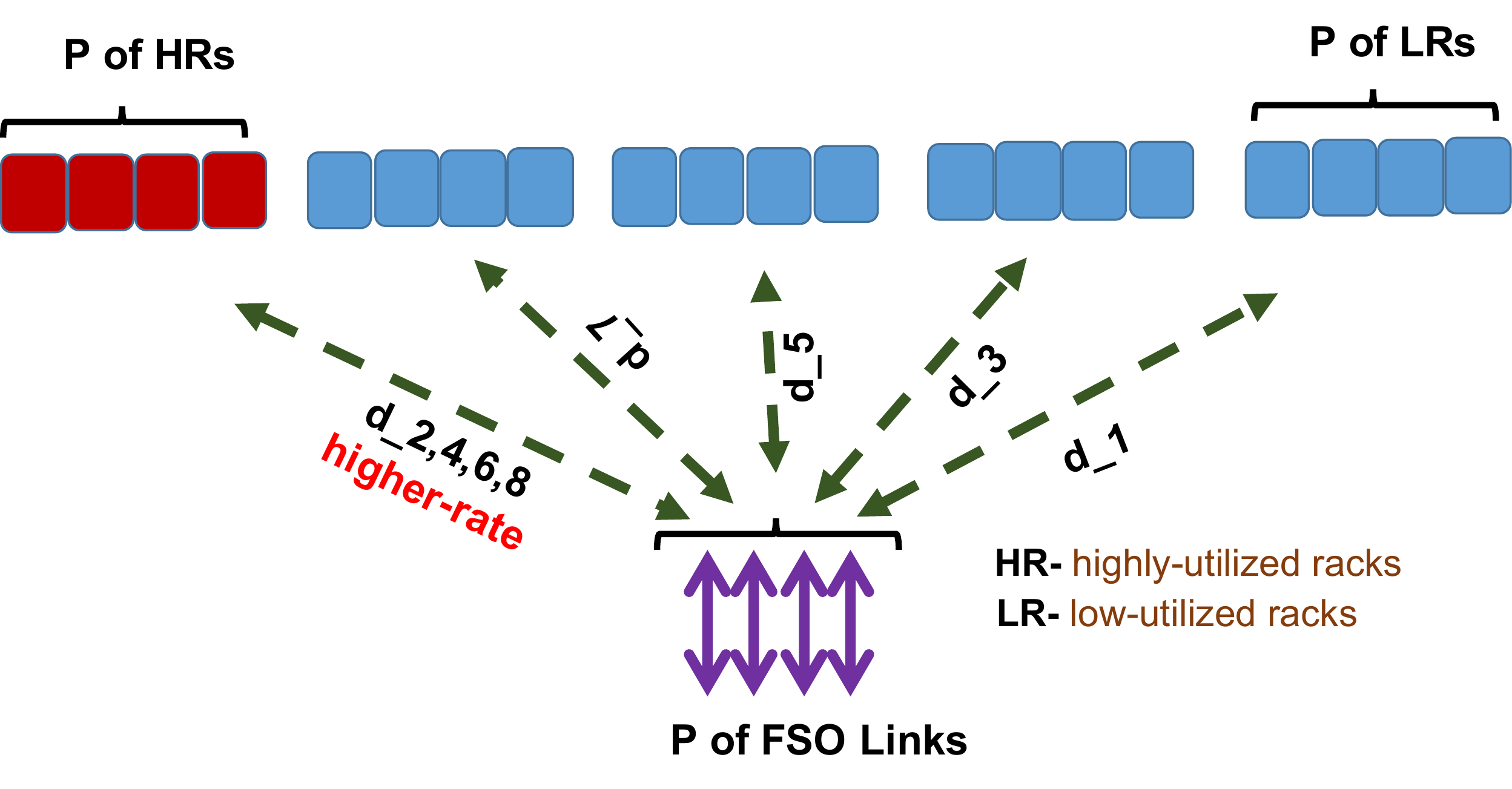}
\caption{Example of FSO bundle scheduling and rotation over $R_{D}$ sets.}
\label{fig:fso-scheduling}
\end{center}
\end{figure}

In order to do this, every beam, in F4Tele, has two mirrors: one horizontal (hm) and another vertical (vm), as shown in Fig.~\ref{fig:microswitch}. The hm is always OFF (in reflecting state), whereas the vm is switchable, similar to the ToR mirrors in Firefly~\cite{firefly}. The vm in the non-transparent "OFF" state is used to create a routing loop between the primary and backup transceivers where the FSO beams are reflected back to the backup transceivers. Fig.~\ref{fig:fso-scheduling} illustrates an explanatory scenario of the introduced solution. The racks in the figure are partitioned into two sets: the set of LRs $R_{1,2,3}$, and the set of HRs $R_{4,5,6}$. The racks of LRs and HRs steer their beams toward different reflectors. When the $d$ of $R_{D}$ set is close to finish, the microswitch changes the state of the corresponding vm to loop the traffic back to the backup transceiver. Instead, we can use a proxy server, one in every  rack, to be in the middle between the local servers and the FSO links. The racks are preconfigured to forward all the control packets toward the local proxy. The proxy keeps the packets in its memory and frequently examines the state of the connectivity with the $R_{M}$ racks and start transmitting the packets when the FSO links are active. The duration of the $d$ and $\tau$ is explained in the next subsection.
\subsection{Rotation Time}\label{sec::rotation}
Our system has P of FSO links rotate together over $R_{D}$s. In every switchover, a new set of $R_{D}$ is connected to the $R_{M}$ for a predetermined period of time, defined here as service-period. The length of the service-period is $d_i$ where $i = 1,2,...,K$ is the index of the $R_{D}$ set that is being connected during $d$ and K is the number of $R_{D}$ sets. Thus, the rotation time $\tau$ = $\sum_{i}^K d_i$. In order to satisfy the  steady-state condition the length of $d$ need to be enough to serve all the waiting packets in the queue under service. In this work the value of $d$ is derived according to the maximum utilization of the racks in the same set. The DC administrator could use historical statistics to measure the value of $d$ for every group~\cite{facebook,projector} and use the programmability of the microswitch and SDN controller to adjust the $d$ length according to the current network statistics.

Without loss of generality, we assume the distribution of arrival-rate is exponential, $\lambda$, where the highly utilized racks $R_{D}^H$ encounter faster arrival-rates, than the least utilized ones $R_{D}^L$. In this work, the $R_{D}$ racks are grouped together in a same set according to their average arrival-rates. Unfortunately, different arrival-rates cause different assignments of service. Such varieties in the service assignments can be provisioned by two methods. First, changing the length of $d$ according the arrival-rate (e.g., long $d$ for $R_{D}^H$). Second, fixing the length of $d$ for all the racks and modifying $\tau$ (i.e., allocating a different number of visits). Therefore, in this work, due to the bursty behavior of DC traffic which produces a sequence of gaps between flowlets and the low percentage of $R_{D}^H$, the second method is adopted, and $d$ length is carefully selected to be flowlet divisible. Fig.~\ref{fig:fso-scheduling} displays an example of our proposed methodology, where the $d$ times are identical. However, the P of FSO links visit the HRs more frequent than the LRs. In the figure all the even slots of $d$, i.e., $d_2,4,6,8,...,$ have been allocated to the HRs whereas the odd slots have been equally divided between LRs, yields every set as a single slot, $\tau_H = d$.

\subsection{Traffic Model}\label{sec:model}
Although the arrival-rate is assumed to be exponentially distributed, the service time has an arbitrary distribution due to the cycles of detaching and attaching the FSO links from rack to another. Such behavior produces arbitrarily distributed vacation intervals. Consequently, we model the system as $M/G/1$ and our model derivations closely follow the standard methods as in~\cite{MG1_1, datanetwork}. We should emphasize that the following mathematical model and analysis can be implemented for both of the highly utilized and least utilized racks.

The FSO link is modeled as a single server facility and is responsible for forwarding the control packets to the $R_{M}$. FSO links are assumed to follow exponentially distributed service times of mean $\overline{\mathcal{X}}_{k}, \: k=1,2,...,K$.. In order to harvest the expected utilities from the control packets and avoid negative sequences from buffering expired control packets, the expected delay experienced by every control packet must not exceed $\mathcal{T}_{\rm{QoS}}$. Therefore, a control packet with a response time longer than $\mathcal{T}_{\rm{QoS}}$ will be dropped. Our objective herein is to compute the impact of the allocated $\tau$ and $d$ on the packet waiting time. Looking into this objective, we aim at studying how the above metrics, particularly the $\tau$ and $d$, vary with FSO graph capacity and dynamics.

\subsection{Delay Analysis}\label{sec:analysis}

As mentioned earlier, the proposed system serves a single set while other sets waiting for their service. The waiting time effects the network performance and the system need to be efficiently optimized to avoid negative consequences and long tail delay. Accordingly in this section we analytically study this delay to hold clear references about the impact of every components that contribute into it. These references help the network engineers in their network design, and understand where to enhance.\footnote{we assume the network designers can add more NMS switches/ports when it is necessary.}. Assuming the network holds $N$ switches and every switch has many ports. Two of them, primary and backup, are dedicated to control packet communications. The FSO link uses the backup port when the primary port gets failed. In our system, the factors that contribute to the communication delay between $R_{D}$ and $R_{M}$ are many. Namely, arrival-rate, the rotation-time, $\tau$, the waiting time in the queue, and the FSO link service time. The service-times of $R_{D}^H$ and $R_{D}^L$ sets' switches are the same. However, their arrival-rates are different, $\lambda_{H}$ and $\lambda_{L}$, respectively. Assuming the  $\lambda_{H}$= $\frac{\lambda_{L}}{\beta}$. In order to fulfill this difference, the $R_{D}^H$ set is served with longer $d$ or visited more frequently which makes $\tau$ value of $R_{D}^H$ shorter, defined herein $\tau_{H}$. The extension in these values, $d$ and $\tau$, should be proportional to the value of $\beta$.

 At first, we are interested in calculating the average waiting time $\overline{\mathcal{W}_{L}^k}$ of set $k$ of $R_{D}^L$. Since the switch specifications and processing capacity in every individual set are assumed to be identical, we look into each set of racks in isolation and going forward the index k will be dropped. Thus, the waiting time of the control packet is given as 
\begin{equation}\label{WHEq0}
\overline{\mathcal{W}_{L}} = \sum_{i=1}^{\mathcal{Z}} \mathcal{X}_{i} + \overline{\mathcal{R}}
\end{equation}
Where $\mathcal{Z}$ is the number of control packets waiting in the queue where the packet arrives. This equation has the control packet, $i$, service-time, $\mathcal{X}_{i}$,and the residual-time $\mathcal{R}$ which is either the residual service-time, $R_{s}$, or the residual vacation-time $R_{v}$. As mentioned earlier, the system has $P$ of FSO channels rotate per $\tau$ to connect the set of $R_{D}$s with $R_{M}$ and this mathematically means the channel service time is extended by an average vacation delay of length $\frac{\tau}{K_{L}}$. Therefore, the first moment waiting time is,   
\begin{equation}\label{WHEq1}
\overline{\mathcal{W}_{L}} = \sum_{i=1}^{\mathcal{Z}} \mathcal{X}_{i} + \sum_{i=1}^{\mathcal{Z}_v} \mathcal{V}_{i} + \overline{\mathcal{R}}
\end{equation}
$\mathcal{Z}_v$ is the number of vacations intervals. Literally, the system under consideration doesn't have an actual vacation. The FSO links serve a set of $R_{D}$s at a time, while other $R_{D}$ sets are waiting (on vacation) for the service. The length of the vacation of a rack is the summation of  $d$s of the switch under service and the subsequent switches until the FSO links serve all the unserved $R_{D}$ sets. The number of sets herein is $K$. 

When a random control packet arrives at the switch, it will wait for the service of all the packets in front of it. Precisely, since each set is allocated a period $d$ and this period has the same length for all the switches in the same group, the packet needs to wait for $d$s in front of it\footnote{In order to simplify the explanations, the $R_{D}^H$ delay model is explained at the end of this section}. Thus,
\begin{equation}\label{WHEq}
\overline{\mathcal{W}_{L}} = \sum_{i=1}^{\mathcal{Z}} \mathcal{X}_{i} + \sum_{i=1}^{\mathcal{Z}_v}d_{i} + \overline{\mathcal{R}}
\end{equation}

The waiting-time model of $R_{D}^H$ set has close similarity with the model of $R_{D}^L$ sets. Like $R_{D}^L$, a newly arriving packet encounters different waiting-times coinciding with the location of arrival and FSO links. The main difference is that the service-period $d$ of $R_{D}^H$s is not enough to serve all the waiting packets in the queue. Hence, a packet may need to wait for multiple $\tau_{H}$s before being served. Also, the $R_{D}^H$ sets are visited more frequently than $R_{D}^L$ sets that enable them to hold a shorter $\tau$ and satisfy the steady-state condition. The Fig.~\ref{fig:fso-scheduling} shows an example where the maximum length of $\tau_{H}$ is one $d$.    

Unfortunately, these characteristics have an apparent influence on the HR waiting-time and complicate the modeling of $\overline{\mathcal{W_H}}$. The main challenge is in the modeling of the number of $d$s the new packet needs to wait. The formal model is finding the probability of every expected state of the new packet and their associate waiting-times. What is the probability that the packet arrives at which $d$ and which rotation, first, second, ... etc. The packet may arrive during $d$, but it would not be served in current service because $d$ is not enough to serve all the packets in front of that new packet. In this case, we have a series of conditional probabilities, conditioned on the arrival location and position of the packet as well as the state of the FSO links. 

The good thing is that the scheduling system assigns the same value of $d$, for every group regardless of their class, $(H,L)$. This feature helps in simplifying the model. This means that the number of $\tau$s the new packet needs to wait or in few words the average waiting time is actually a summation of multiple $d$s where the value of $d$ is a known constant. Assuming the number of $d$s the packet needs to wait is $Z_{d}$. Then, 
\begin{equation}\label{WHEq}
\overline{\mathcal{W}_{H}} = \sum_{i=1}^{Z_d} d_{i} +\overline{ \mathcal{R}}.
\end{equation}
Where the $Z_{d}$=$K_{H}+K_{L}$ and because the scheduling algorithm visits only one $R_{D}^L$ set after serving $R_{D}^H$ set. Hence, $K_{L}$=1, and $Z_{d}$=$K_{H}+1$. In order to understand the waiting-time distribution, the model needs to consider the state-probabilities of every expected waiting-times. The state-probability is the probability that one of  $R_{D}$ set is under service when the packet arrives. The significance of this probability manifested because the present location of the FSO links, (i.e., which $R_{D}$ set is under service), determines how many $d$s the new arrivals need to wait for. For example, if packet $i$ arrives at an $R_{D}$ set while the FSO links have just left, the packet needs to wait the full rotation time, $\tau_{max}$ which is much longer than if it arrives at an under-service switch. Thereby, the waiting-time equation needs to cover all possible lengths of $\tau$, where the $\tau$ length is bounded down by a single $d$ and up by a full rotation, $\tau_{max}=\sum_{i}^I d_i$, including $d$s of $R_{D}^H$ sets. We should emphasize that due to the scheduling algorithm the waiting-time of $R_{D}^L$ sets is longer than and differs from the waiting-time of $R_{D}^H$ set. The scheduling algorithm herein visits the $R_{D}^H$ more frequently to accommodate its high arrival-rate. Thus, when a packets arrives at one of $R_{D}^L$ sets its waiting-time is,
\begin{equation}\label{GeneW}
\mathcal{W}_L = Pr_1\cdot D_1 + Pr_2\cdot D_2 + \cdot\cdot\cdot Pr_{k-1}\cdot D_{k-1} +Pr_h\cdot D_h +\overline{ \mathcal{R}}
\end{equation}
\begin{align}
D_1 = \sum_{i=2}^{K} d_{i}.
, D_2 = \sum_{i=3}^{K} d_{i}, D_3 = \sum_{i=4}^{K} d_{i}, ...
\end{align}
\begin{equation}\label{CondEq}
D_h = \sum_{i=1}^{K-1} P(R_{D}^{K-i}|R_{D}^{K-1-i})\cdot Pr_{K-1-i} \cdot \overline{D_{k-i}}.
\end{equation}
The consideration of the probability $Pr_{i}$ and the waiting-time of every state, generalize the above mathematical model. This generalization makes the model suitable for general cases such as when the arrival-rate of the LR sets is unequal, which is out of the scope of this study. In this work the bundle of FSO links has been scheduled to serve both $R_{D}^L$ and $R_{D}^H$ sets, where the set of $R_{D}^H$ is served more frequent than $R_{D}^L$. When a random control packet arrives at one of $R_{D}^L$ set needs to wait for the FSO bundle to finish serving all the sets in the front including the $R_{D}^H$ set. As stated in equation (\ref{GeneW}) the $D$s is the expected waiting-times and $Pr$s are their probabilities.

However, the FSO bundle scheduling divergence arises when the $R_{D}^H$ is under service, rather than selecting the $R_{D}^H$ after every $R_{D}^L$ set, like the above example, the micro-switch selects afterward set according to the preceding $R_{D}^L$ set. Hence, we use the conditional probability $P(R_{D}^{k-i}|R_{D}^{k-1-i})\cdot Pr_{k-1-i}$ conditioning on the probability of preceding state. The conditional probability is necessary in calculating $D_h$ which is the expected waiting time when the $R_{D}^H$ is under service where the packet is waiting at one of the $R_{D}^L$ sets.

The $Pr_{i}$ is the state-probability that the switches of set $i$ is under service, where $i$ could be any set from $R_{D}^L$. In the same context, the $Pr_h$ is the state-probability of $R_{D}^H$ set. In order to find $Pr_{i}$, we need to understand the service scheduling procedure of the presented solution. In this work, the FSO link scheduling follows an unconventional scheduling procedure, where every $R_{D}^L$ set is served after serving one of the $R_{D}^H$ sets, here we have a single set. This infers that the system can be easily divided into two general groups: $R_{D}^L$ and $R_{D}^H$ sets, where the FSO bundle alternates between them. The FSO bundle either serves the $R_{D}^L$ group or $R_{D}^H$ group, where the $R_{D}^H$ herein consists only of a single set, its probability $Pr_h$ is 1/2. Moreover, the scheduling algorithm handles the probabilities of the $R_{D}^L$ sets equally. Therefore,
\begin{equation}\label{WHEq}
Pr_1=Pr_2=Pr_3 = \cdot\cdot\cdot = Pr_{k-1}.
\end{equation}
\begin{equation}\label{WHEq}
Pr_1+Pr_2+Pr_3 + \cdot\cdot\cdot + Pr_{k-1} = (1-Pr_h).
\end{equation}
Therefore,
\begin{equation}\label{WHEq}
Pr_i = \frac{(1-Pr_h)}{K_L},  \forall i \in \mathcal{K_L}.
\end{equation}
The waiting time formula of $R_{D}^L$ is, 
\begin{dmath}\label{WLEq}
\mathcal{W}_{L} = \rho_L\cdot \Big[\sum_{k=1}^{Z_i}\overline{\mathcal{X}_k} \Big]+(1-\rho_L)\cdot \Big[Pr_i\cdot [\frac{(K-1)(K-2)d}{2}] + Pr_h \cdot D_h\Big]+\overline{ \mathcal{R}}
\end{dmath}
The waiting time of $R_{D}^H$ is, 
\begin{equation}\label{WHEq}
\mathcal{W}_{H} = \rho_H\cdot \Big[\sum_{k=1}^{Z_i}\overline{\mathcal{X}_k} \Big]+(1-\rho_H)\cdot \Big[\sum_{k=1}^{Z_i}\overline{\mathcal{X}_k} + \sum_{j=1}^{R_{D}^p} d_j \Big]
\end{equation}
Where $Z_i=\lambda_{i}\overline{W_i}$, $R_{D}^p$ is the number of non-hotspot data racks that need to be served after serving a single hotspot rack, herein is one, and $\rho$ is the probability a packet arrives during a service\cite{datanetwork}.

In order to complete our analysis we need to find the first moment and second moment of $\mathcal{R}$. The first moment of the residual time is defined as
\begin{equation}\label{MeanREq}
\overline{ \mathcal{R}} = \frac{\overline{1}}{2}\left( \frac{\rho \cdot\overline{ \mathcal{X}^2}}{\overline{ \mathcal{X}}}\right).
\end{equation}
where $\text{Var}(\mathcal{R}) = \overline{\mathcal{R}^2} - \overline{\mathcal{R}}^2$, and 
\begin{equation}
\overline {\mathcal{R}^2} =\frac{1}{3} \left(\lambda\overline{\mathcal{X}^3} \right).
\end{equation}
\begin{figure}[t!]
    \centering
        \includegraphics[width=0.7\columnwidth]{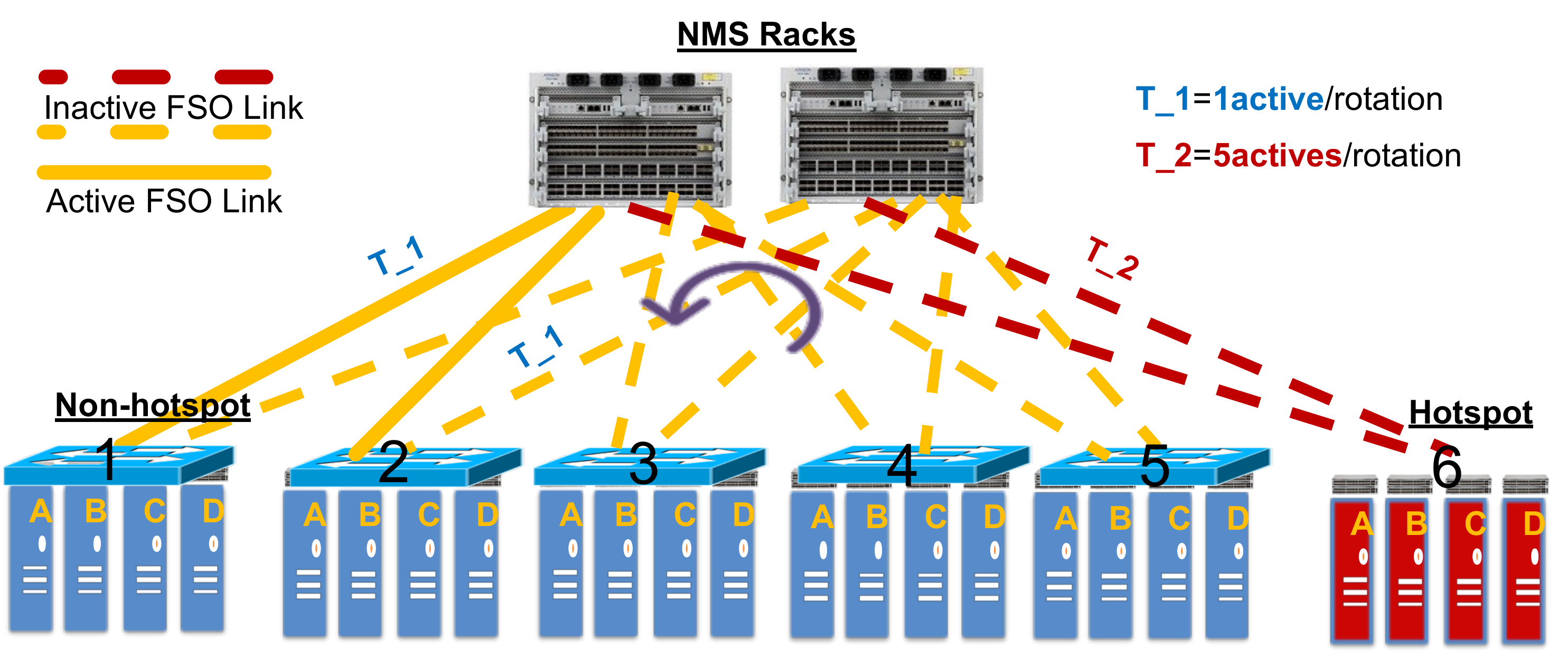}
        \caption{Network topology used for evaluation.}
        \label{fig::eva}
\end{figure} 
\section{Implementation and Evaluation}\label{sec:evaluation}  
In this section, we present our implementation as well as evaluation setup and scenarios. We aim at studying the impact of FSO link rotation and $d$s on the $R_{D}$ to $R_{M}$ communications. In order to fulfill this, we divided the evaluation into two cases according to the value of the service period. In the first case, $d$ is 10 milliseconds, and we used 100 milliseconds during the evaluation of the second case. In every individual case, we studied the proposed solution by using two types of traffic (TCP and UDP), separately. Besides that, we divided the racks into two parts (hotspot and non-hotspot) according to the flow arrival-rates. The hotspot traffic sources which occupy about 10\% of the total traffic sources, are configured with a high arrival-rate, which is 10x higher than the non-hotspot traffic sources. This skewed arrival-rate enables us to use a non-sequential rotation procedure, as explained in Fig.~\ref{fig:fso-scheduling}. 

In this rotation procedure, the hotspot rack is served immediately after serving a single non-hotspot rack that enables the hotspot racks to secure more frequent $d$s; which means less rotation-time. However, this procedure complicates our evaluation task and the collection of output results. In a few words, the evaluation has been turned from a single sequential long procedure into multiple parallel small procedures. In every small procedure, the traffic generation function of a single rack is activated. The time needed by a traffic generation function to trigger all its flows is longer than the service period, and we need to comply with this period to avoid serving a rack for longer than its $d$. In order to tackle this challenge, we use a multi-threading technique whereas every thread is responsible for the traffic generation of a single rack and continue the execution of its flow even after its service-period is expired. The thread of the second rack is immediately triggered after the $d$ of the current rack is expired regardless of whether it initiated all of its flows or not. We should emphasize that some of the flows in every rack start during the vacation-time.  
\subsection{Network Setup}
In this work, we conducted our evaluation by using Mininet emulator~\cite{mininet} to obtain an evaluation environment close to a real data center network. Mininet uses the underline system resources and operating system to build the configured network topologies including its switches, hosts, and links. The hosts are real virtual hosts, and the switches are real software switches, i.e., OVS switches~\cite{openvswitch}. We use Mininet to build the topology, as shown in Fig.~\ref{fig::eva}, which has 4 NMS racks and 20 data racks, each with 5 hosts/servers.\footnote{We select this network size because we assume the network designers can increase the number of FSO links by adding more NMS switches/ports.} The data racks are divided into 5 non-hotspot sets and one hotspot set, each set has 4 racks which means 4 FSO connections rotate together to serve these sets. The sets from one to five are the non-hotspot sources, and set 6 is the hotspot source. The flows are generated by Iperf and arrive according to an exponential distribution with different means in symmetry with the non-hotspot and hotspot sets which are 20 millisecond and 2 milliseconds, respectively. In order to diversify our evaluation, the size of every flow was randomly selected from a range of 1MB to 10MB for both of the TCP and UDP traffics. The link speed is 1Gbps where we expected the speed is much higher than this in real DCs.
\subsection{Algorithms}
\textit{Benchmark} is the expensive and complex solution where it has a direct link from every data rack to the NMS racks. The links do not interfere with each other and work almost in an isolated environment to enable them to secure the maximum possible efficiency.
\textit{F4tele} is the proposed solution where a bundle of FSO links serves a single group of data racks for a period of time (10 ms, or 100ms), then jumps to serve another group. In this solution rather than dropping the traffic of unserved groups, it is looped back to go through the secondary FSO transceiver exploiting its buffer capacity.
 \textit{F4tele+} is F4tele without looping back the traffic of unserved set of racks from primary to backup interface.   
\subsection{TCP 10ms and 100ms Results}
\begin{figure*}[t!]
     \centering
      \begin{subfigure}[b]{0.45\linewidth}  
        \includegraphics[width=\columnwidth]{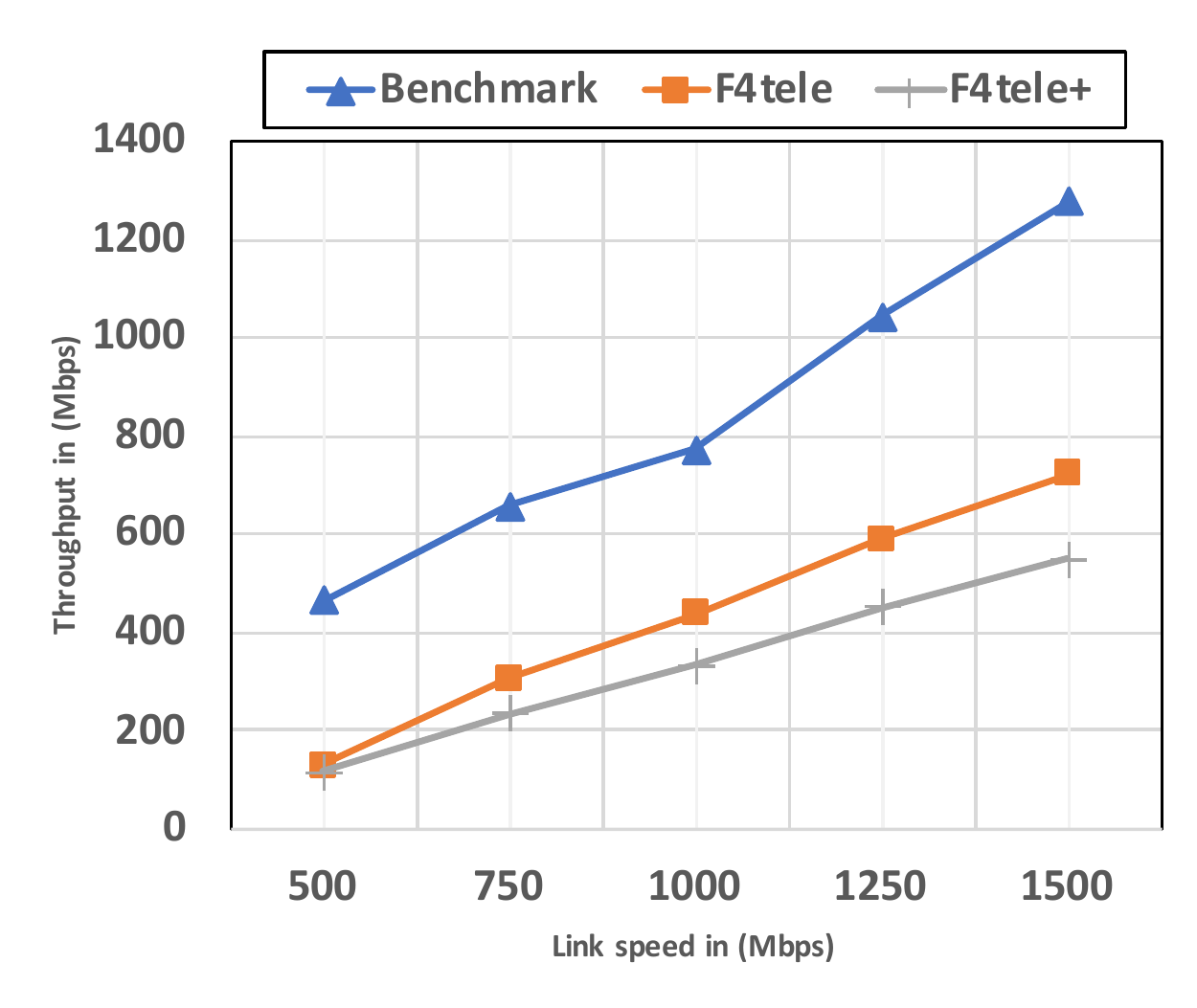}
        \caption{Non-hotspot racks during 10 milliseconds $d$.}
        \label{fig::tcp10}
    \end{subfigure}
      \begin{subfigure}[b]{0.45\linewidth}  
        \includegraphics[width=\columnwidth]{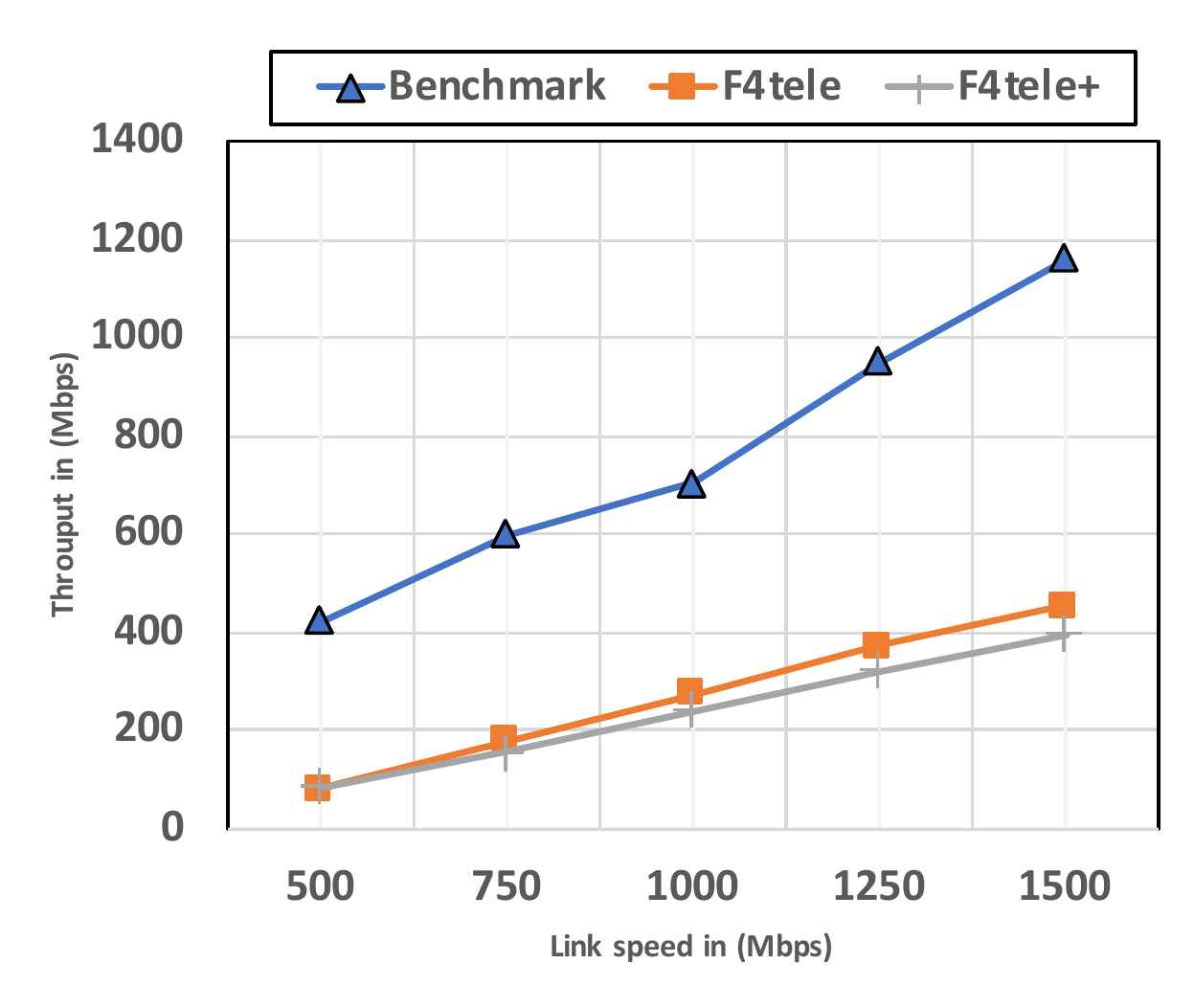}
        \caption{Hotspot racks during 10 milliseconds $d$.}
        \label{fig::tcp10hot}
     \end{subfigure}
      \begin{subfigure}[b]{0.45\linewidth}  
        \includegraphics[width=\columnwidth]{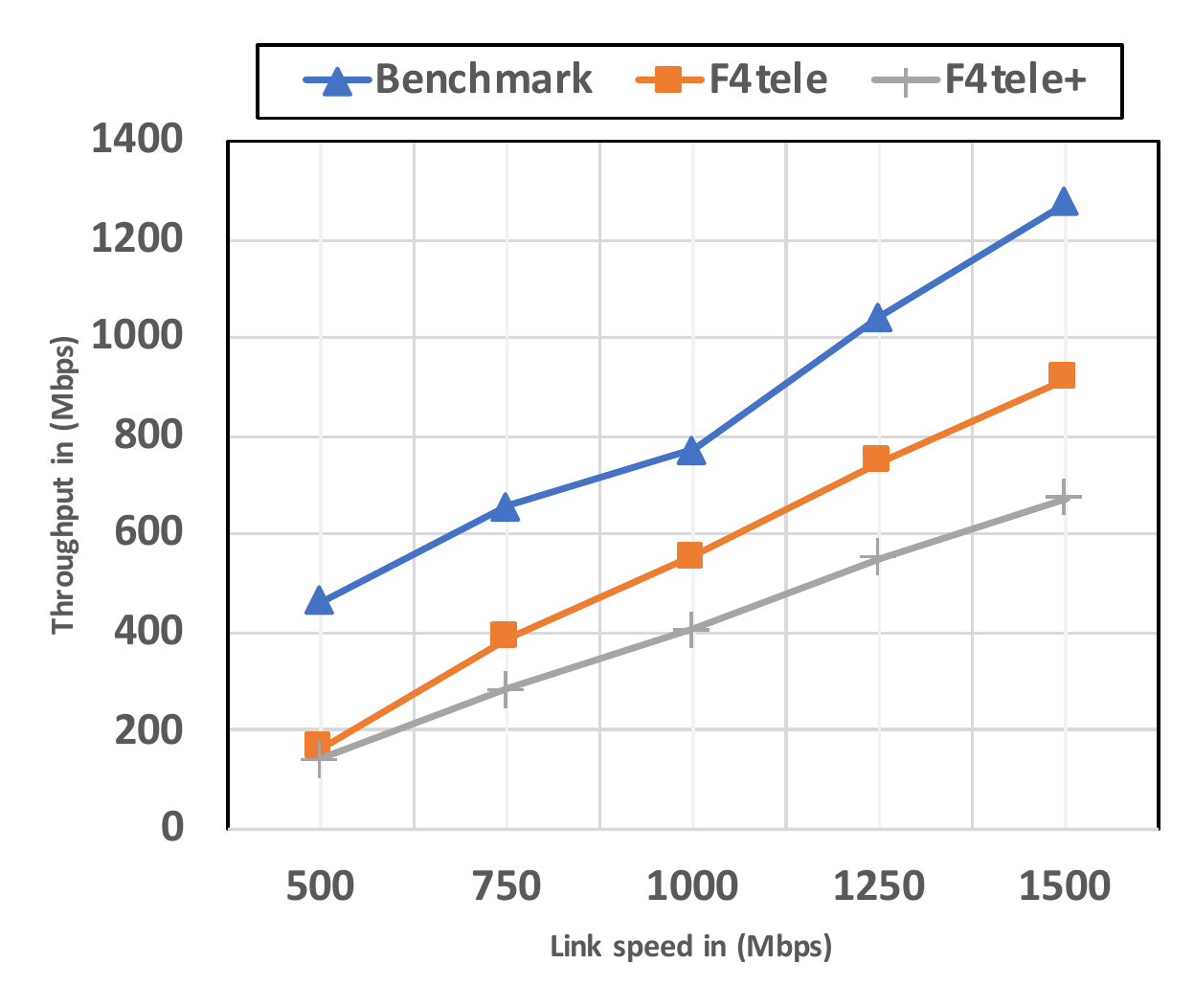}
        \caption{Non-hotspot racks during 100 milliseconds $d$.}
        \label{fig::tcp100}
     \end{subfigure}         
      \begin{subfigure}[b]{0.45\linewidth}  
        \includegraphics[width=\columnwidth]{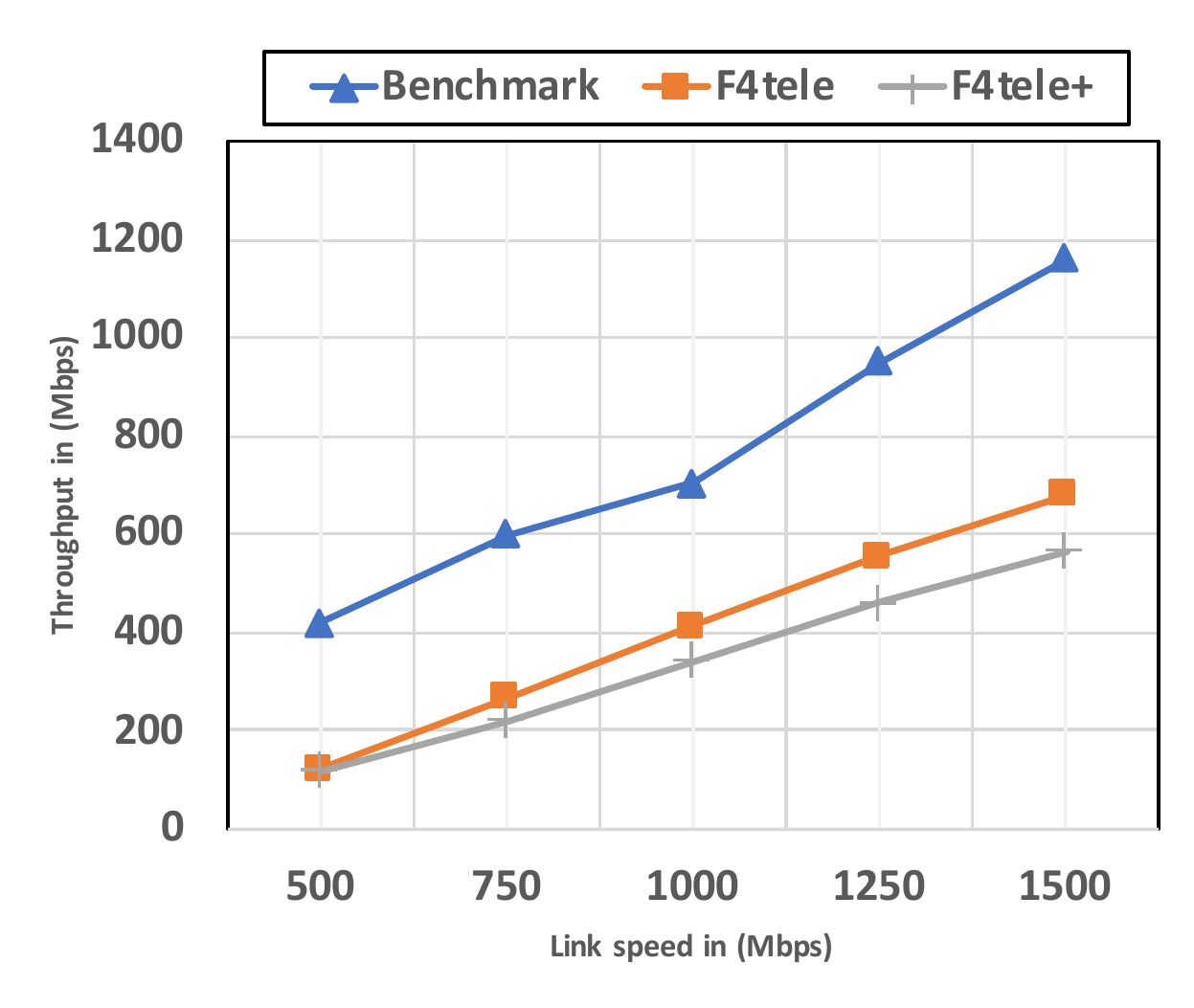}
        \caption{Hotspot racks during 100 milliseconds $d$.}
        \label{fig::tcp100hot}
             \end{subfigure}         
                 \caption{Average throughput of non-hotspot and hotspot racks for TCP flows during 10 and 100 milliseconds $d$.}
\end{figure*}

In this subsection we present the performance figures of the proposed solution with and without the packet looping technique compared with the benchmark.  In this part the length of $d$ is (10, and 100 ms) and Iperf is configured with TCP protocol. The configurations are identical for all the evaluated solutions. The results of non-hotspot flows are separated from the results of hotspot flows where their results are illustrated in Fig.~\ref{fig::tcp10hot} and Fig.~\ref{fig::tcp100hot}, and Fig~\ref{fig::tcp10} and Fig.~\ref{fig::tcp100}, respectively. Normally, the benchmark solution always shows the highest throughput among other solutions and this due to the establishment of a direct FSO link from every data rack to the management racks. Although this feature isn't provisioned to the F4Tele its average throughput for non-hotspot flows is about 450Mbps which is about 60\% of the benchmark throughput. Likewise, the throughput of F4Tele+ is 334Mbps which is about 55\% of the benchmark, as shown in Fig~\ref{fig::tcp10} and Fig.~\ref{fig::tcp10hot}.  

\begin{figure}[t!]
    \centering
        \includegraphics[width=0.7\columnwidth]{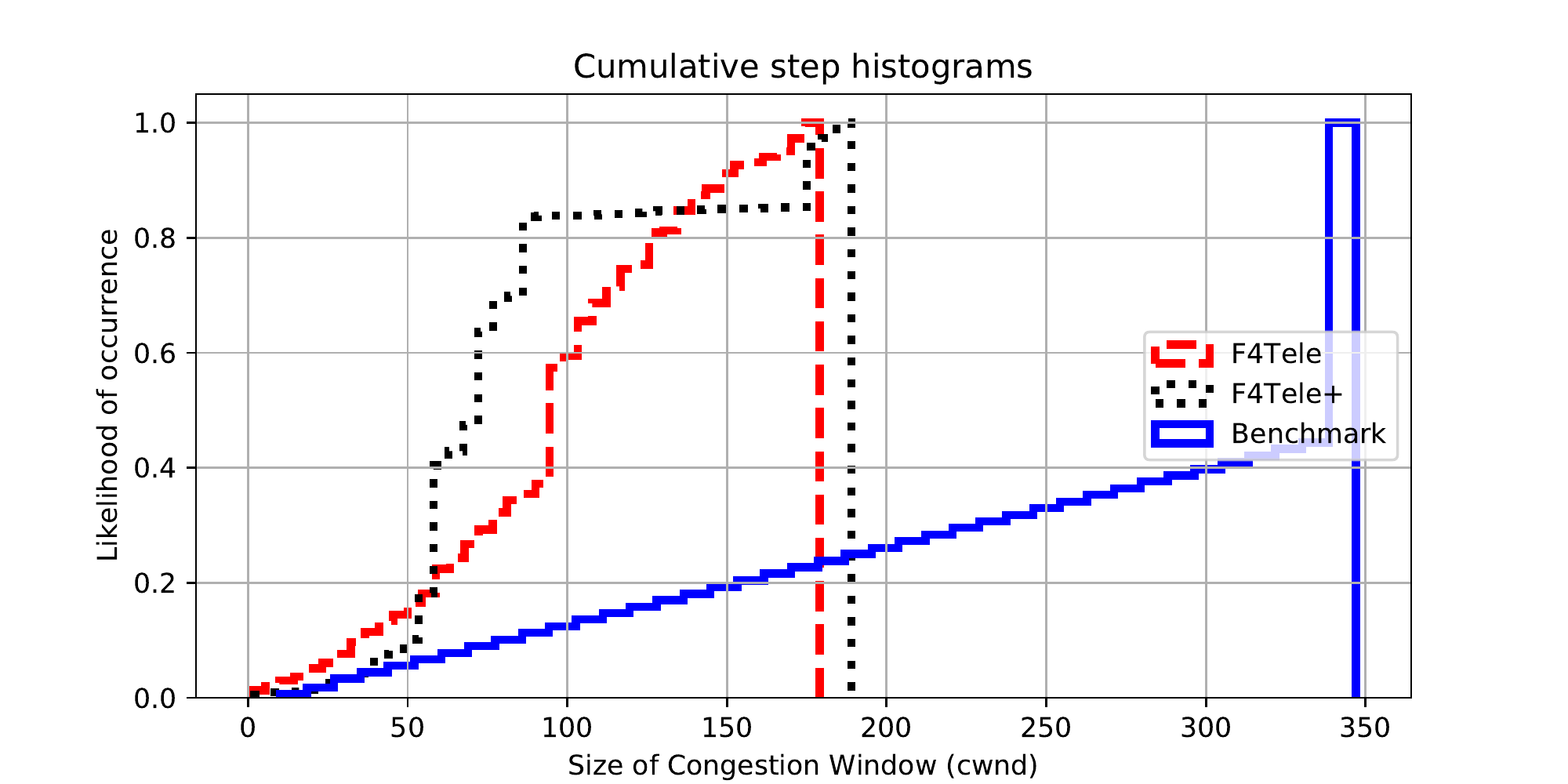}
        \caption{CWND CDF during 10 milliseconds $d$.}
        \label{fig::cwnd10}
\end{figure}

\begin{figure}[t!]
    \centering
        \includegraphics[width=0.7\columnwidth]{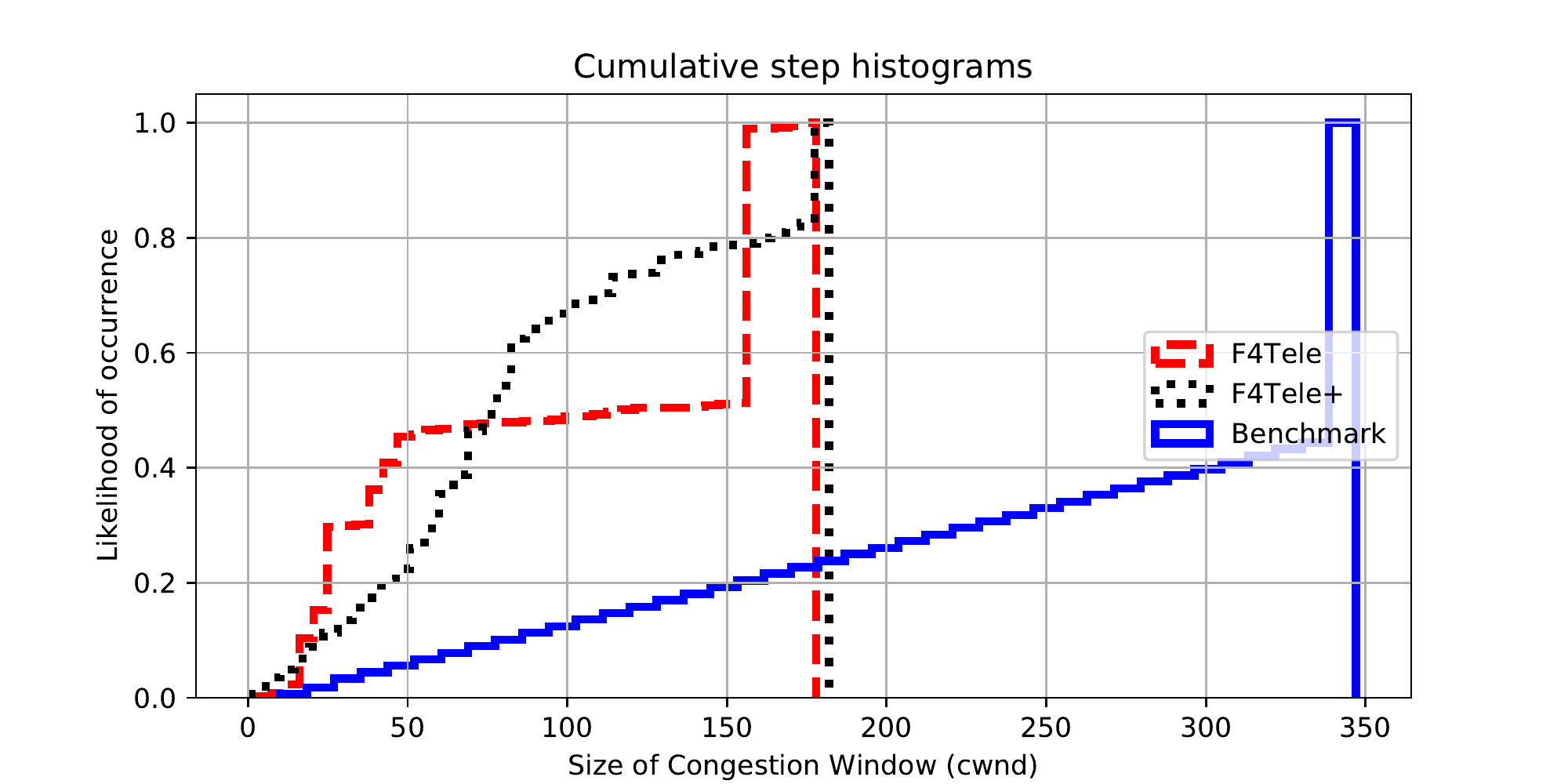}
        \caption{CWND CDF during 100 milliseconds $d$.}
        \label{fig::cwnd100}
\end{figure}

When we increase $d$ to 100ms, the TCP congestion control scheme finds enough space to efficiently utilize the link capacity. Particularly, the congestion-window (cwnd) has enough time to enlarge its size and approach higher link utilization. Consequently, the throughput of both versions of F4Tele has been increased as well. The F4Tele achieves 556MBps average throughput which is about 25\% lower than the benchmark, while the throughput of F4Tele+ is 410Mbps which is 45\% lower than the benchmark, as displayed in Fig.~\ref{fig::tcp100}. In hotspot, both of F4Tele versions encountered throughput degradation due to the high flow arrival-rate, as displayed in Fig.~\ref{fig::tcp100hot}. This decay is more clearer in the 10ms period results because the 10ms is not long enough to support the cwnd to recover from the dropping and approach large size and maximize the utilization. This phenomenon is clearly demonstrated in Fig.~\ref{fig::cwnd10} and Fig.~\ref{fig::cwnd100}.  The cwnd CDF of F4Tele and F4Tele+ is almost similar during 10ms evaluation. However, the cwnd CDF results of both solutions are enhanced during 100ms evaluation because the cwnd has longer time to grow and maximize the link utilization.   
     

\begin{figure*}[t!]
    \centering
      \begin{subfigure}[b]{0.45\linewidth}  
        \includegraphics[width=\columnwidth]{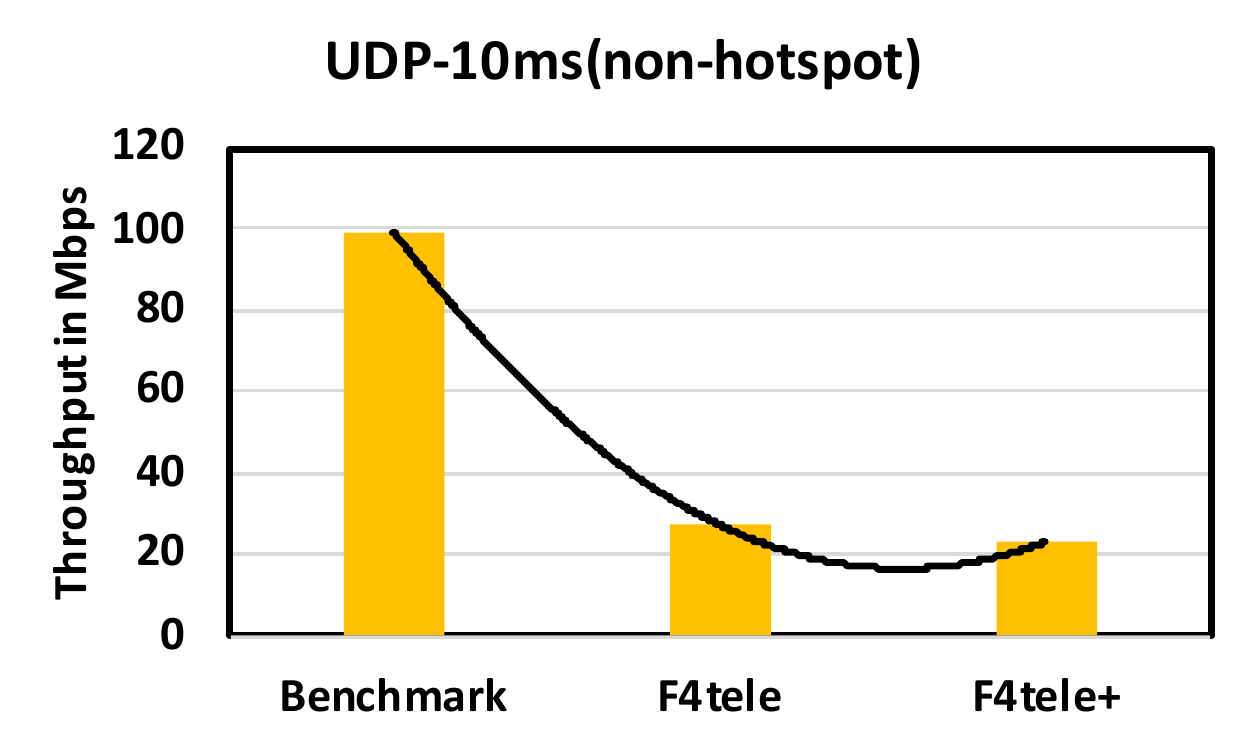}
        \caption{Non-hotspot racks during 10 milliseconds $d$.}
        \label{fig::udp10}
    \end{subfigure}
      \begin{subfigure}[b]{0.45\linewidth}  
        \includegraphics[width=\columnwidth]{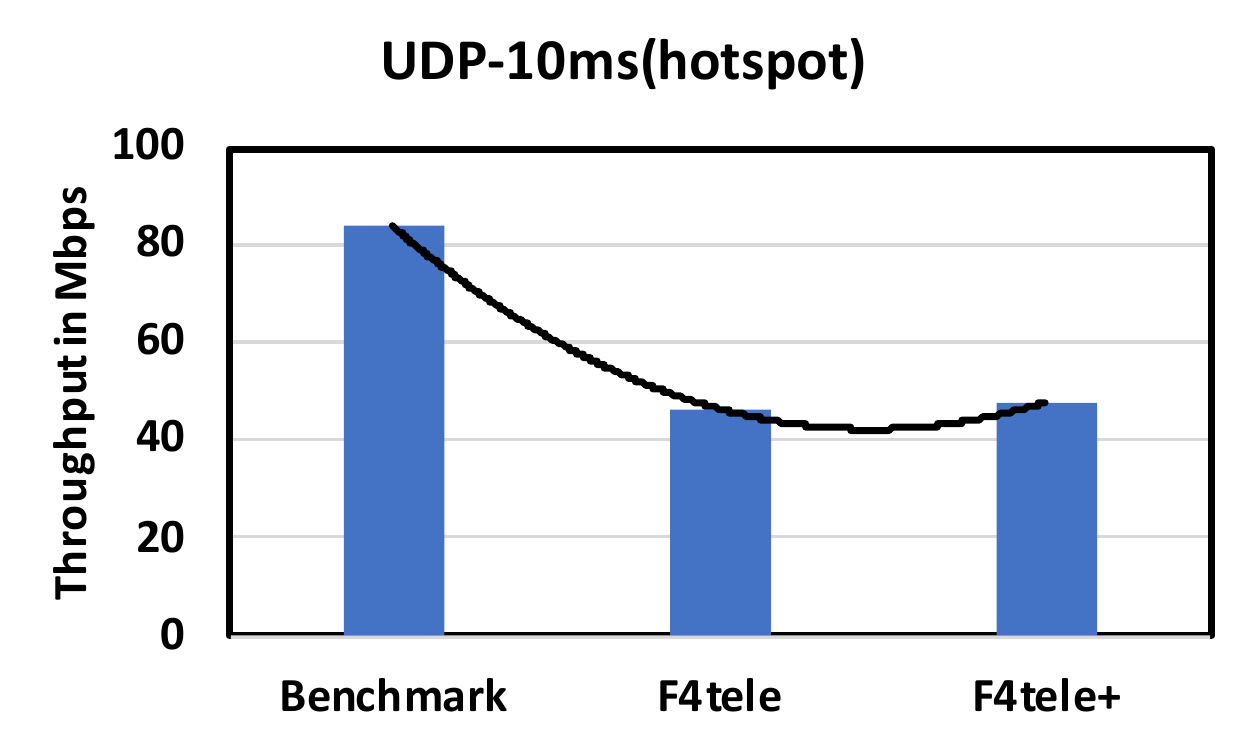}
        \caption{Hotspot racks during 10 milliseconds $d$.}
        \label{fig::udp10hot}
     \end{subfigure}
      \begin{subfigure}[b]{0.45\linewidth}  
        \includegraphics[width=\columnwidth]{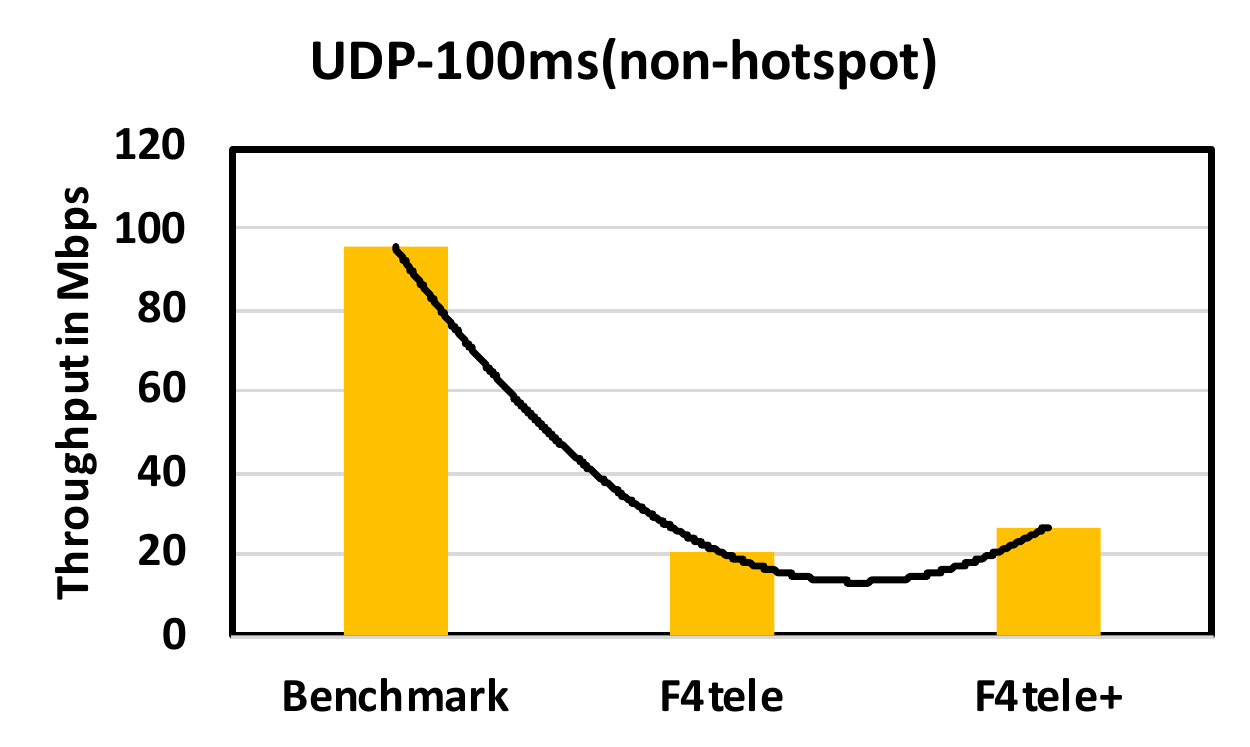}
        \caption{Non-hotspot racks during 100 milliseconds $d$.}
        \label{fig::udp100}
     \end{subfigure}         
      \begin{subfigure}[b]{0.45\linewidth}  
        \includegraphics[width=\columnwidth]{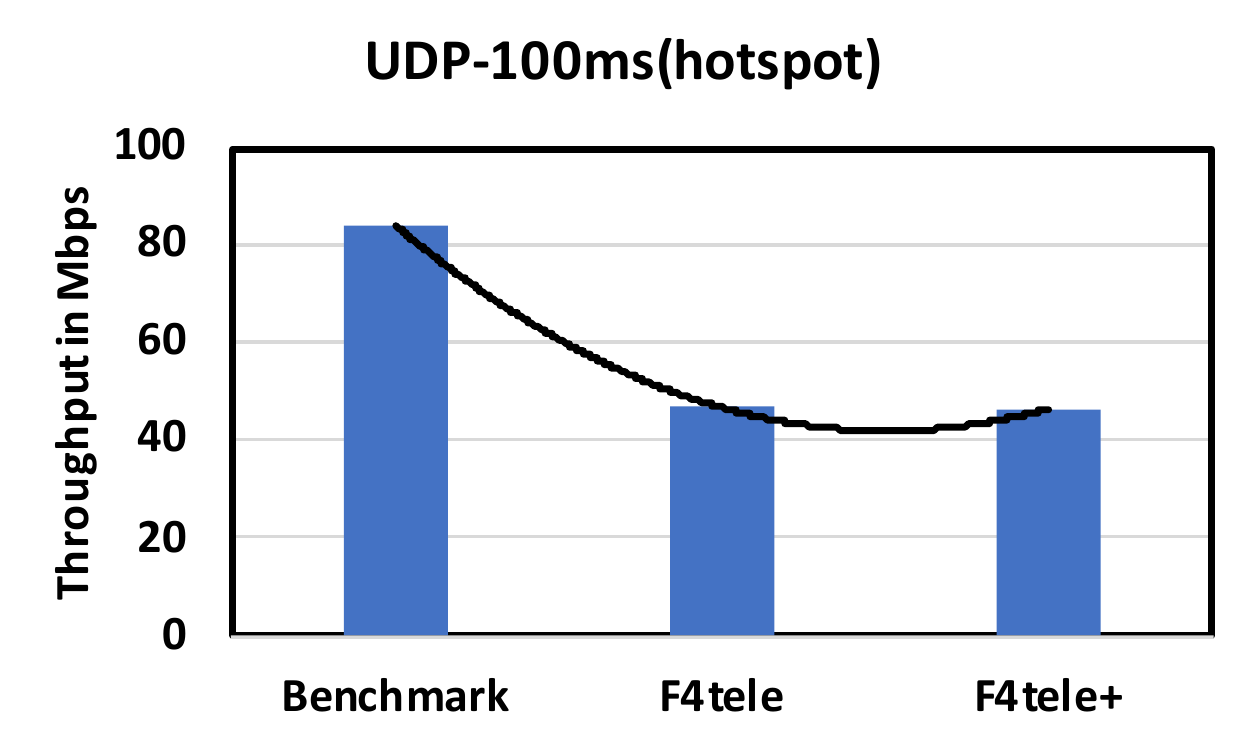}
        \caption{Hotspot racks during 100 milliseconds $d$.}
        \label{fig::udp100hot}
             \end{subfigure}         
                 \caption{Average throughput of non-hotspot and hotspot racks for UDP flows during 10 and 100 milliseconds $d$.}
\end{figure*} 
\subsection{UDP-10ms and 100ms Results}
The evaluation wouldn't be complete if we don't study the impact of the introduced solutions on UDP flows. Unlike TCP, the UDP protocol doesn't have congestion or flow control mechanisms. The sources keep sending the packets regardless to the status of the network or destinations. We find that the average throughput of UDP flows increases with the increase in the number of serving times. In contrast, the previous results showed that the TCP throughput increases with the length of $d$. This behavior is mainly because the TCP sets its parameters, such as the cwnd size, according to the network status and needs enough time to reach the optimal settings. However, the UDP is opportunistic and doesn't react to the network status or change its parameters, accordingly. 

In this part of evaluation we use the same configurations and settings of the previous part. However, Iperf is configured with UDP protocol and with transmitting bandwidth is 100Mbps. The configurations are identical for all the evaluated solutions. Similarly, the results of non-hotspot flows are separated from the results of hotspot flows where their results are illustrated in Fig.~\ref{fig::udp10} and Fig~\ref{fig::udp100hot}, respectively. The F4Tele achieves average throughput for non-hotspot flows about 28Mbps which is  about 30\% of the benchmark. Likewise, the throughput of F4Tele+ is 24Mbps which is about 25\% of the benchmark. Unlike TCP, when $d$ is increased to 100ms, we didn't encounter a noticeable variations in the flow throughput.  When $d$=10ms, the F4Tele achieves average throughput for hotspot flows about  47Mbps which is  56\% of the benchmark. While, with $d$=100ms, the average throughput is  47.1Mbps. Likewise, the throughput of F4Tele+ is 48Mbps which is about 57\% of the benchmark and its average throughput is almost the same, when $d$=100ms. Moreover, sometimes, such as non-hotspot(100ms) and hotspot(10ms), the F4Tele is better than F4Tele. This means that the looping technique isn't beneficial for UDP flows which is expected because the UDP destination doesn't have a packet reordering scheme. In general, the UDP destination consumes the arrived packets regardless to their order. As a result, in our final system we suggest to add a filter in the looping scheme to filter out the UDP traffic from the looping packets. 
\begin{figure}[t!]
    \centering
        \includegraphics[width=0.7\columnwidth]{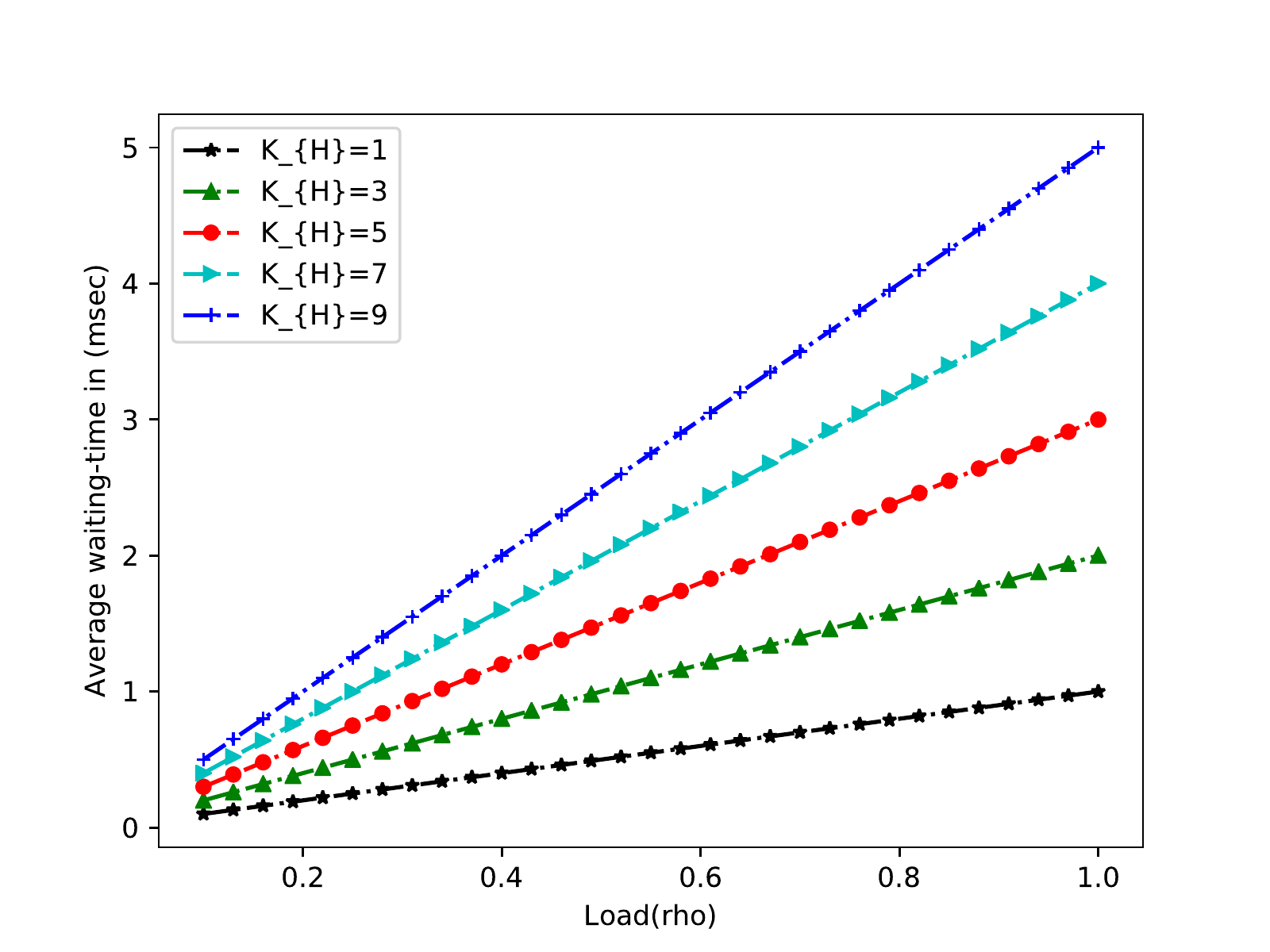}
        \caption{Average waiting-time for hotspot sets when $d$ equals 10 milliseconds, and different number of hotspot sets.}
        \label{wh10}
\end{figure}   
\begin{figure}[t!]
    \centering
        \includegraphics[width=0.7\columnwidth]{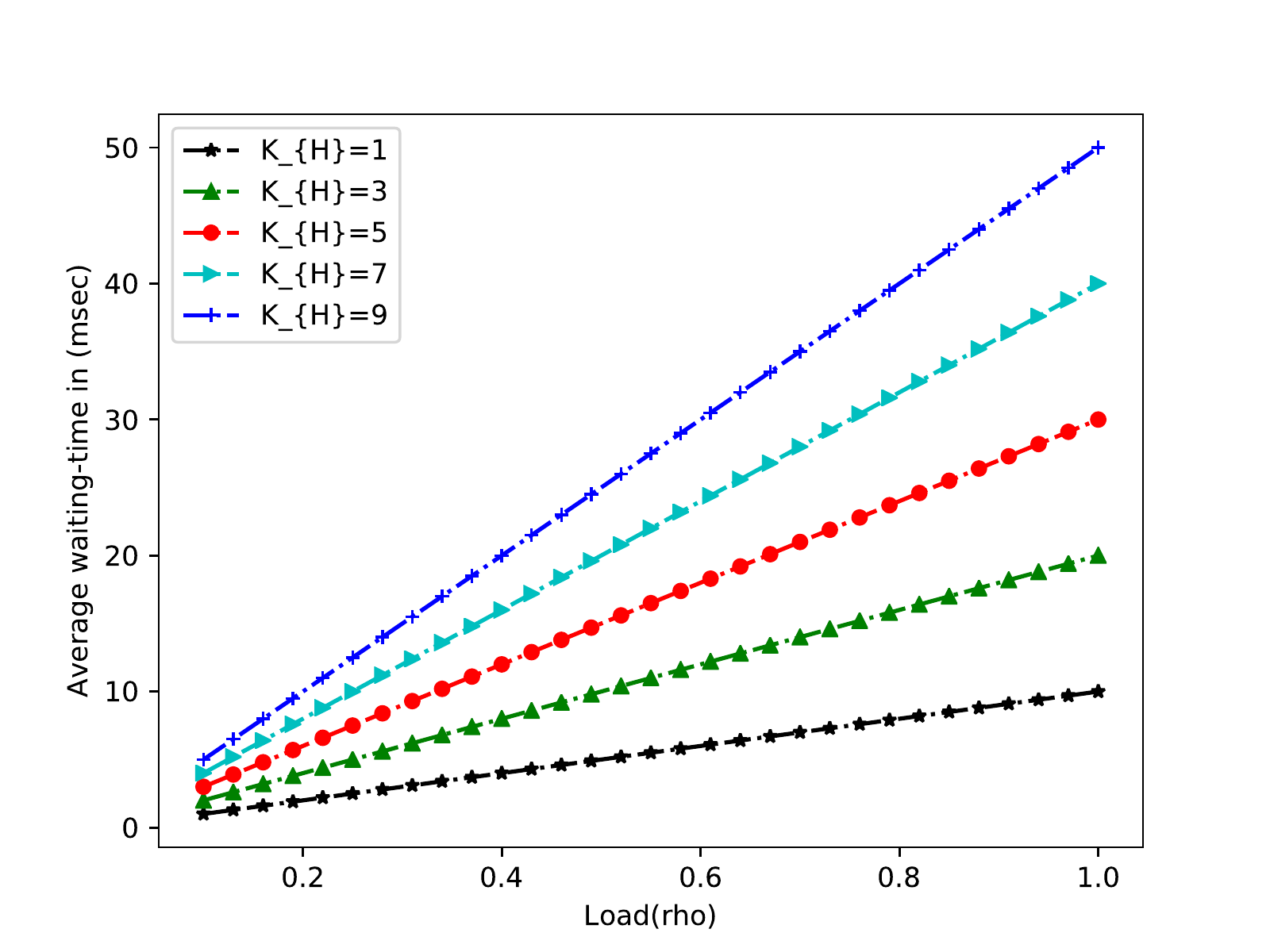}
        \caption{Average waiting-time for hotspot sets when $d$ equals 100 milliseconds, and different number of hotspot sets.}
        \label{wh100}
\end{figure} 
\begin{figure}[t!]
    \centering
        \includegraphics[width=0.7\columnwidth]{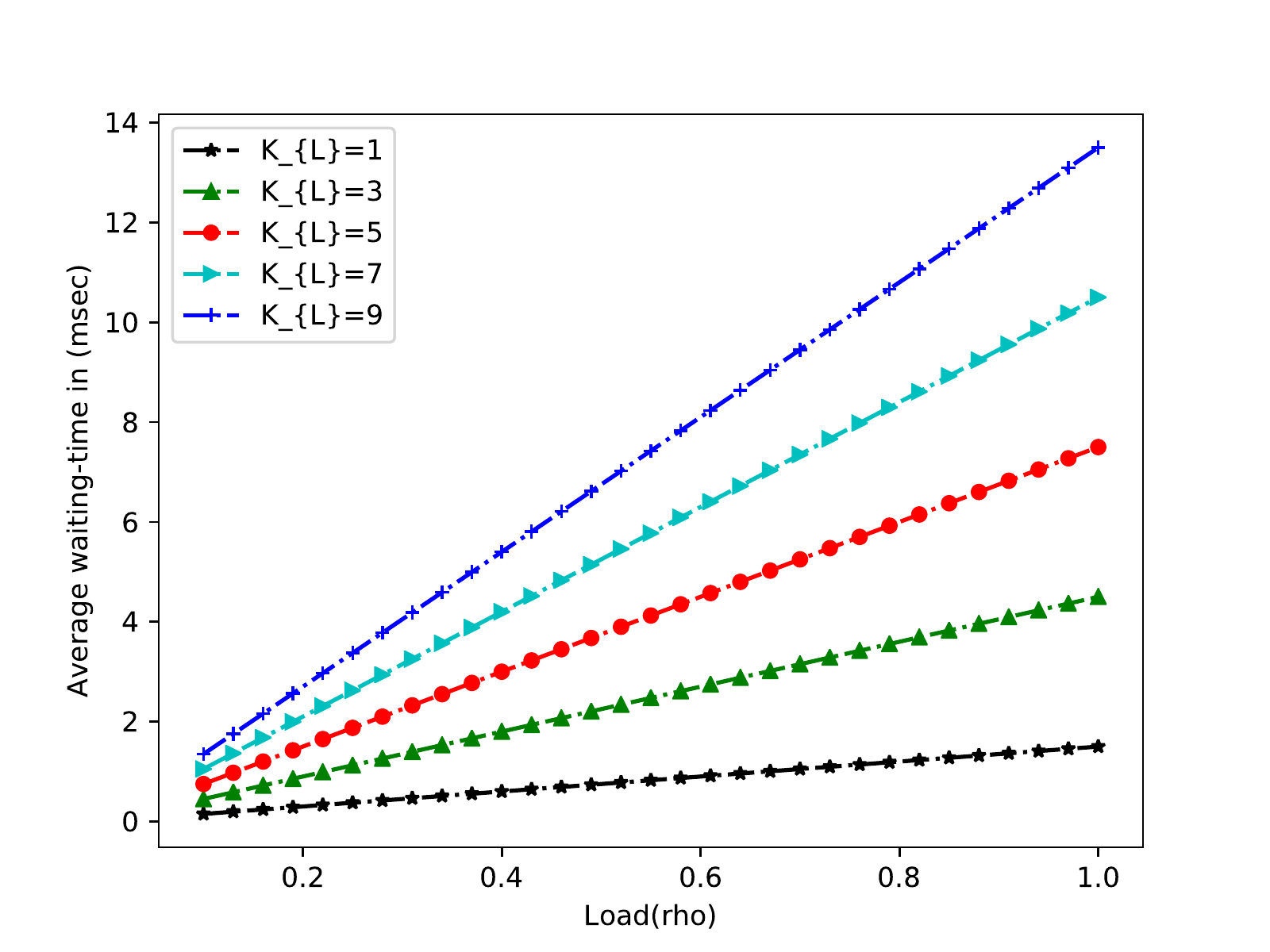}
        \caption{Average waiting-time for non- hotspot sets when $d$ equals 10 milliseconds, and different number of non-hotspot sets.}
        \label{wl10}
\end{figure}   
\begin{figure}[t!]
    \centering
        \includegraphics[width=0.7\columnwidth]{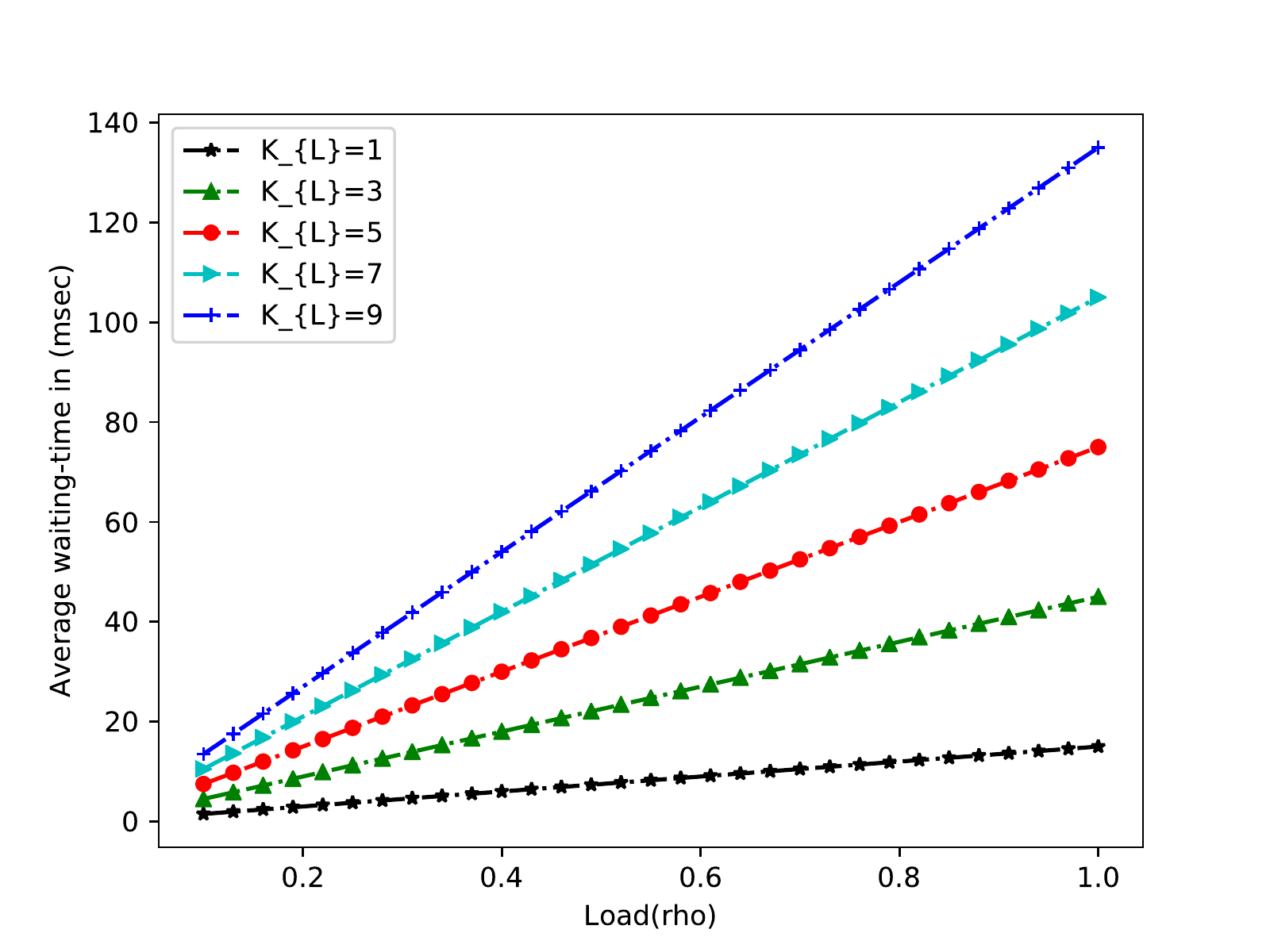}
        \caption{Average waiting-time for non-hotspot sets when $d$ equals 100 milliseconds, and different number of non-hotspot sets.}
        \label{wl100}
\end{figure}  
\begin{figure}[t!]
    \centering
        \includegraphics[width=0.7\columnwidth]{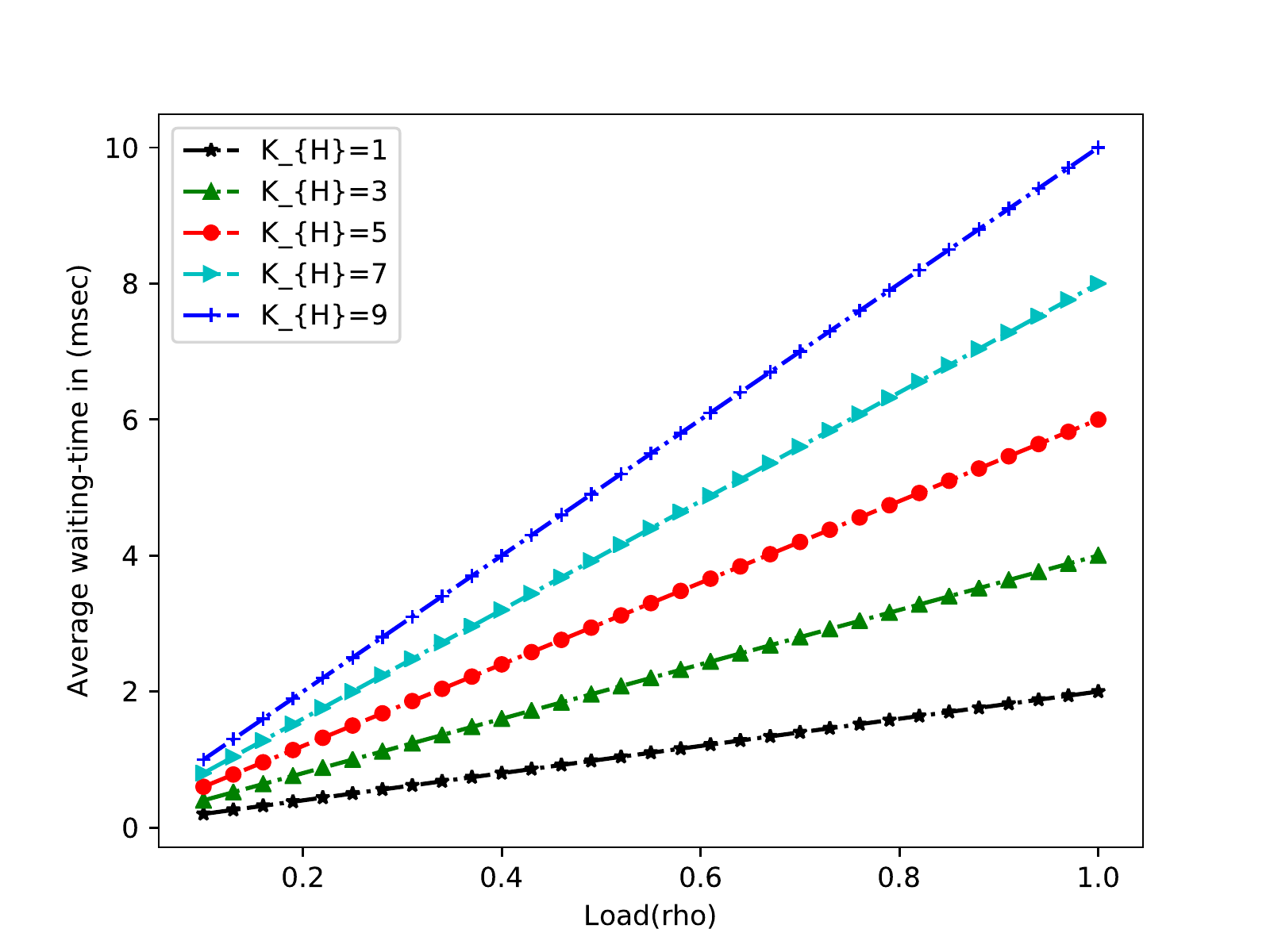}
        \caption{Average waiting-time for hotspot sets when $d$ equals 10 milliseconds and with two times slower FSO speeds.}
        \label{wh10n10}
\end{figure}   
\begin{figure}[t!]
    \centering
        \includegraphics[width=0.7\columnwidth]{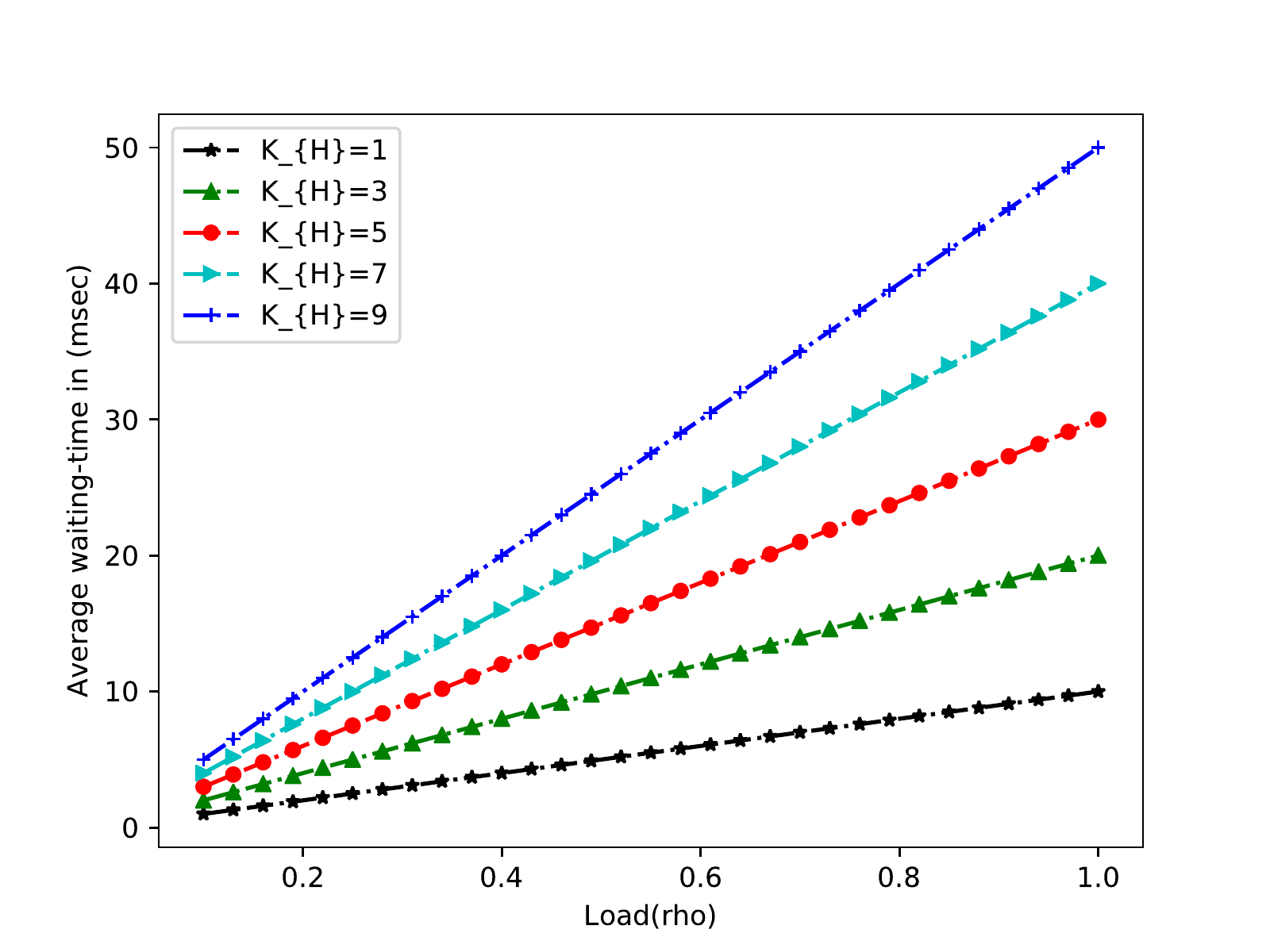}
        \caption{Average waiting-time for hotspot sets when $d$ equals 10 milliseconds and with ten times slower FSO speeds.}
        \label{wh10n2}
\end{figure} 
\subsection{Delay Results}
There are some factors contribute to the waiting-time of NMS traffic, namely length of $d$, speed of FSO links, number of sets, workload, and the response time of the resource allocation algorithm. The evaluation of these factors help the system designer to select the appropriate settings that match the network quality of service requirements. In this subsection we study the impact of these factors on the F4Tele performance. The F4Tele system is similar to other systems where some of its factors such as the traffic workload, are uncontrollable and the designer needs to cohabit with. However, other factors such as the value of $d$, the number of FSO links and their capacities are freely to be tuned and adjusted to match the system performance needs. Every individual factor has distinct influence on the overall system performance. In this part of evaluation, we study the impact of these factors individually. The system delay has been studied with $d$ equals 10 and 100 milliseconds, and during different FSO link capacities (service-time $\mu$). For the other parameters (i.e., utilization and number of data racks) we use a complete range of values; 10-100\% for the load and 1 to 10 for the number of set of data racks. The average-waiting-time results for hotspot set is shown in Fig.~\ref{wh10}, and Fig.~\ref{wh100}, when the value of $d$ is 10 and 1oo milliseconds, respectively. The figures show five lines, one line per the change on the number of $R_{D}^H$ sets. This to show the relation between the waiting-time and the number of $R_{D}^H$ sets. Thus, the waiting-time positively increases with the increase in the load and the number of $R_{D}^H$ sets. Similarly, the results of non-hotspot sets demonstrate the same trend behavior of the hotspot set. The results of non-hotspot sets are displayed in Fig.~\ref{wl10}, and Fig.~\ref{wl100}. 

In order to study the impact of another contributor on the average waiting-time (i.e., FSO link speeds $\mu$) the delay of hotspot set has been re-evaluated with slower FSO link speeds. Fig.~\ref{wh10n2}, and Fig.~\ref{wh10n10}, portrayed the results when the FSO speed is reduced by ten times, and two times, respectively.  The impact is clearly appear in the increase of the waiting-time.


\section{Conclusions}\label{sec::conclusion}
Network management applications seek for large volume of packets to execute fine-grained analysis. Unfortunately, capturing and forwarding large volumes of traffic cripples data networks or degrades their performance. In general, to mitigate this challenge, existing network management frameworks either recraft network switches to perform some of the network management functions, or adopt an expensive method that builds a dedicated network for the captures, defined herein as control-packets, or reduce the number of captures on account of some necessary debugging features. 

In this work, rather than forwarding control-packets through data networks or providing coarse-grain information on network status, we explored a novel free-space optics (FSO) based scheme to interconnect data racks with the racks of network management servers (NMS). Our approach enables all data racks to establish direct FSO links with NMS racks to carry the control traffic. FSO technology has several merits: re-configurable topology, extremely high-speed links, low cabling complexity and low maintenance challenges, as well as inexpensive appliances. Unfortunately, pointing FSO links from every data rack toward the NMS racks is practically impossible unless we build an expensive and completely dedicated network. Alternatively, we leveraged the FSO agility to develop a rapid topology reconfiguration and routing method without transceiver alignment challenges.  In the evaluation F4Tele achieved average throughput for non-hotspot flows about 560Mbps which is 72\% of the benchmark throughput. Likewise, the throughput of F4Tele+ is 410Mbps which is about 55\% of the benchmark. In brief, we found that the average throughput of UDP flows increases with the increase in the number of serving times. In contrast, the previous results showed that the TCP throughput increases with the length of the service-period.          

\bibliographystyle{IEEEtran}
\bibliography{Reffull}
\end{document}